\documentclass[structabstract]{aa}

\usepackage{graphicx}
\begin{document}

\title{Spectral ageing analysis and dynamical analysis of the double-double radio galaxy J1548$-$3216}

\author{J. Machalski\inst{1},  
M. Jamrozy\inst{1}, \and  C. Konar\inst{2}}

\institute{Astronomical Observatory, Jagellonian University,
ul. Orla 171, PL-30244 Krakow, Poland,
\email{machalsk@oa.uj.edu.pl, jamrozy@oa.uj.edu.pl}
\and
Indian Institute of Astrophysics, Block II, Koramangala, Bangalore - 560 034, India\\
\email{chiranjib.konar@gmail.com}
}
\date{Received , 2009; accepted , 2009}

\abstract
{ Determining ages of the outer and the inner lobes of so-called double-double radio
galaxies (DDRGs) is crucial for understanding the active cycles of galactic nuclei,
the phases of interruption of the jet flow, and physical conditions in the
surrounding galactic and intergalactic medium governing the jets' propagation. A
recognition and understanding of these conditions during the restarted jet activity
is of special interest.}
{We determine the ages and other physical characteristics of the outer and the inner
lobes of the DDRG J1548$-$3216, as well as the properties of the surrounding
environment during the original and the restarted phase of the jets' activity.}
{Using the new low-frequency and high-frequency radio images of this galaxy, we
determined the shape of the spectrum along its lobes and performed the classical
spectral-ageing analysis. On the other hand, we applied the analytical model of the
jet's dynamics, which allowed us to derive the physical conditions for the source's
evolution during the original jet propagation through the unperturbed IGM, as well as
those when the restarted new jet propagates inside the outer cocoon formed by the old
jet material that passed through the jet terminal shock.}
{The dynamical age estimate of the outer and the inner lobes is 132$\pm$28 Myr and
$\sim$9$\pm$4 Myr, respectively. The synchrotron age in the outer lobes
systematically rises from $\sim$25 Myr in the vicinity of the lobes' edges to about
65--75 Myr in the centre of the old cocoon. These ages imply an average expansion
speed along the jets' axis: (0.012$\pm$0.003)$c$ in the outer lobes and (0.058$\pm
$0.025)$c$ in the inner lobes, but the latter speed would be $\sim$0.25\,$c$ when
they were of age less than 1 Myr. We find that the jet power during the restarted
activity is about ten-fold fainter than that of the original jet. Similar
disproportion is found for the internal pressures and the magnetic field strengths
in the old cocoon and those in the inner lobes. This disproportion can
be effectively reduced by assuming the same equations of state for the emitting
particles and the magnetic fields within the old and the new lobes. However,
we think that our assumption of the non-relativistic equation of state for the old
cocoon and the relativistic one for the new lobes is more justified.}
{}

\authorrunning{J. Machalski et al.}
\titlerunning{Spectral and dynamical ages of GRG J1548$-$3216}
\maketitle

\section{Introduction}

Although the intrinsic time evolution of powerful radio sources of Fanaroff-Riley
type II (FRII; Fanaroff \& Riley 1974) is largely understood and described with a
number of analytical models (e.g. Kaiser, Dennett-Thorpe \& Alexander 1997; Blundell,
Rawlings \& Willott 1999; Manolakou \& Kirk 2002; Kino \& Kawakatu 2005), there are
still several unanswered questions about the duty-period of the active galactic
nucleus (AGN), the jet production processes, its interaction with the external
gaseous environment including the intergalactic medium (IGM), and the contents of the radio lobes as a part of the low-density `cocoon'.   

The double-double radio galaxies (DDRGs) are characterized by two pairs of 
unequally-sized edge-brightened (FRII-type) lobes sharing the same radio core. In
most of them the outer and inner double structures are aligned well. The existence of
such radio sources is the evidence that the jet activity in AGN may be not continuous
during the lifetime of a source. In fact, an intermittent production of jets can be
 connected with stochastic transitions between two accretion modes: the standard one
-- with angular momentum transmitted outwards by viscous torques within the accretion
disk -- and the `magnetic' one, with the developed large-scale magnetic fields and
related MHD winds (Nipoti, Blundell \& Binney 2005; K\"{o}rding, Jester \& Fender
2006). Sikora, Stawarz \& Lasota (2007) incorporated the above idea into the spin
paradigm scenario. Postulating that the efficient production of relativistic jet
requires both a large black hole (BH) spin (as in the model of Blandford \& Znajek
1977) and an efficient collimation mechanism (cf. Begelman \& Li 1994), they noted
that the intermittent jet activity observed in active galaxies accreting at high
rates may be due to intermittent collimation of the central Poynting flux-dominated
(so called `Blandford \& Znajek') outflow by heavier and slower MHD wind generated in
the inner parts of the accretion disk undergoing state transitions. In the
framework of this interpretation, the jet axis in the subsequent jet activity epochs
is expected to be the same, since this axis is determined by the spin of the central
BH, which should not change substantially on short ($\ll 100$ Myr) timescales (see
the discussion in Sikora et al. 2007).

Such an interrupted production of jets is evidently imprinted in the radio morphology
of DDRGs. We are interested in certain aspects of these sources. Are the ages and
internal densities of the inner lobes much lower than these values for most `normal'
radio sources of similar physical size as the inner lobes? Is the density of the 
pre-existing cocoon much lower than in the unperturbed galactic and IGM environment,
or is this density higher? The first case would strongly suggest that the new inner
structure is formed in a channel drilled through the old cocoon by the former jet
activity cycle, which has been modelled by the numerical MHD simulations of Clarke \&
Burns (1991). They predict that `the restarted jet will always be overdense (denser
than its immediate surroundings) if the original jet is underdense relative to the
quiescent IGM'. While the restarting jet model accounts for many of the observations,
there remain some profound discrepancies difficult to be reconciled (cf. Clarke
1997). The second case would imply an efficient replacement of the inner lobes by the
heavier external medium (e.g. Kaiser, Schoenmakers \& R\"{o}ttgering 2000; Brocksopp
et al. 2007).

The radio galaxy J1548$-$3216 (PKS B1545$-$321) is a remarkable example of DDRG in
which the newly restarted jets propagate through the remnant cocoon of a previous
active phase (Subrahmanyan, Saripalli \& Hunstead 1996; Saripalli, Subrahmanyan \&
Shankar 2003, hereafter referred to as SSS2003). This galaxy has recently been extensively studied by Safouris et al. (2008, hereafter referred to as S2008), especially under the aspect of an observational constraint for the  2D and 3D numerical simulations of the restarted jet provided by Clarke \& Burns (1991) and Clarke (1997), respectively. In S2008, the authors suggest that observational data are consistent with a picture that the restarted jets generate narrow-bow shocks, and the inner lobes in this galaxy are a mixture of cocoon plasma re-accelerated at the bow shock and new jet material re-accelerated at the termination shock. They propose that the evolution of the restarted jets and the inner lobes is strongly influenced by an entrainment of the external IGM into the pre-existing cocoon. 

In this paper the spectral-ageing and dynamical analyses of J1548$-$3216 are
performed with the aim of (i) determining the synchrotron age distribution in the outer lobes of this galaxy (in the old cocoon), (ii) estimating of this age in the inner lobes, (iii) comparing these ages with the dynamical ages estimated with the DYNAGE algorithm of Machalski et al. (2007), (iv) determining the jet powers during the first and the second phase of activity, as well as other physical parameters characterizing the lobes and their environments, such as the particle density, energy density, internal pressure, magnetic field strength and its density, propagation speeds
of the lobes along the jets' axis, etc. Most working approaches in this paper are similar to those applied in our previous publications on other giant-sized radio galaxies: DDRG J1453+3308 (Konar et al. 2006) and a further ten selected galaxies (Jamrozy et al. 2008; Machalski, Jamrozy \& Saikia 2009, hereafter referred to as MJS2009). The analyses presented in this paper are based on the new radio observations recently conducted with the Giant Metrewave Radio Telescope (GMRT) and Very Large Array (VLA), and on the Australia Telescope Compact Array (ATCA) and VLA archival data kindly provided to us by Vicky Safouris and Ravi Subrahmanyan.  The new observations and the data reduction are presented in Sect.~2. The resulting total-intensity 334, 619, 1384, 2495, and 4860 MHz total-intensity images are used in Sect.~3 to derive radio maps of the outer double structure of the investigated galaxy, as well as to extract the inner double structure from a background of the underlying cocoon formed during the earlier phase of the nuclear activity. The spectral-ageing analysis of the outer and the inner structures is described in Sect.~4. while the dynamical analysis is presented in Sect.~5. Results of these analyses, as well as our contribution to the aspects of the restarted nuclear activity, the environmental conditions ruling the new jets propagation within the relict cocoon, and their energetics -- derived with another
approach than applied in the previous studies of this radio galaxy by SSS2003 and S2008 -- are discussed in Sect.~6.

For the purpose of calculating the linear size, volume, and luminosity of the lobes,
we use cosmological parameters $\Omega _{\rm m}$=0.27, $\Omega_{\rm vac}$=0.73, and
$H_{0}$=71 km\,s$^{-1}$Mpc$^{-1}$.

\section{Observational data and their reduction} 

The observing log for all the observations is listed in Table~1, which is arranged as
follows. Columns 1 and 2 show the name of the telescope and the array configuration
for the former and the recent VLA observations; columns 3 and 4 give the frequency of
observations and the primary beamwidth; columns 5 and 6 show a typical angular
resolution and an rms noise level achieved in the resulting images of the radio
galaxy investigated. The last column 7 gives the dates of the observations. More
details of these observations are given below.

\begin{table*}[t]
\centering
\caption{Observing log.}
\begin{tabular}{llccccl}
\hline
\hline
Teles- & Array & Obs.   & Primary & Ang.  & rms   & Observing \\
cope   &       & freq.  & beam    & resol.  & noise & date      \\
       &       & [MHz]  & [arc\,min]&[arc\,sec] & [mJy\,beam$^{-1}$]\\

\hline
GMRT   &       &$\;\:$334 & 80      &$\:$15   & 0.24  & 2008, May 24 \\
GMRT   &       &$\;\:$619 & 43      & 7.5     & 0.12  & 2008, Mar 8 \\
ATCA+VLA* &    & 1384     & 40      & 3.5     & 0.04  & 2001, Dec 2 \\
VLA*   & BnA   & 1384     & 30      & 3.5     & 0.04  & 2002, May 31 \\
ATCA*  &       & 2495     & 24      &$\;\:$5  & 0.05  & 1999, Mar 7 \\
VLA    & DnC   & 4860     &$\;$9    &$\:$15   & 0.03  & 2008, Jun 5 \\
VLA*   & BnA   & 4910     &$\;$9    & 1.3     & 0.02  & 2002, Jun 2 \\
\hline
\end{tabular}

$^{*}$: Archival data; courtesy of Vicky Safouris and Ravi Subrahmanyan. 
\end{table*}

The low-frequency GMRT observations at 334 and 619 MHz were made in the standard
manner, with each observation of the target source interspersed with observations of
calibrator sources. The phase calibrators B1714$-$252 (at 334 MHz) and B1626$-$298
(at 619 MHz) were observed after each of several 20 min-lasting exposures of the
target centred on the core position. 3C286 was used as the flux density and bandpass
calibrator based on the scale of Baars et al. (1977). At each of the two frequencies
the total observing time on the target source was only about 150 min because of very
limited observing time scheduled for the project. Unfortunately, a large part of
334-MHz data were strongly affected by radio frequency interference, and these data
had to be flagged in the reduction process, which further reduced the quality of the
data. Acceptable data were edited and reduced with the NRAO {\sc AIPS}
package. All these data were self-calibrated to produce the best possible images.

At the frequencies of 1384 and 2495 MHz, the archival data taken with the ATCA and
VLA arrays are used. In particular, the ATCA 2495 MHz map of the total structure
published by SSS2003, as well as the combined ATCA+VLA 1384 MHz of the total structure
and the high-resolution VLA 1384 and 4910 MHz map of the inner double published by
S2008. For the purpose of specifying a high-frequency spectrum of diffuse lobes of
the outer double structure, we made other 4860 MHz observations of the target source
with the VLA in its DnC configuration. Again, 3C286 and B1522$-$275 were used for the
amplitude and the phase calibrations, respectively. Two 20 min exposures of the
fields centred on each of the two outer lobes were reduced, self-calibrated, and
combined into one image of the entire source.

\section{Observational results}

\subsection{New radio images}

A full-resolution GMRT 334 and 619 MHz images, as well as the VLA/DnC 4860 MHz image,
are presented in Fig. 1. Our new images, especially those at low frequencies,
confirm the overall morphology of J1548$-$3216 already presented and discussed by
SSS2003 and S2008, i.e. that both the outer lobes are edge-brightened and rather
sharply bounded. Likewise in those papers, our images also do not show any evident
hot spots or very compact structures at the ends of the lobes, and both low-frequency
images confirm a distinct pair of emission peaks along a bright rim at the western end
of the NW lobe.  As in the archival ATCA and ATCA+VLA data, the inner double structure
(a pair of relatively bright, narrow lobes) is strongly immersed into the diffuse
bridge of emission extended from the bright edges of the outer lobes towards the
radio core. A flare of the bridge transverse to the source's axis in the vicinity of the core, very well shown at the ATCA 1384 MHz image in SSS2003, is also pronounced at
both low frequencies. This flare is missing at the 4860 MHz image, which suggests a
very steep radio spectrum in that part of the structure (cf. Sect.~3.4). However, a
missing flux density at this frequency is negligible. Indeed, the area marked with
the dashed line in the right panel of Fig.\,1 is about 60 restoring beams, and a
missing flux is likely between the rms noise level and the first contour C1 in this
image, both multiplied by $\sim$60 beams, i.e. between $\sim$1.8 mJy and $\sim$6 mJy.
Even the missing flux of 6 mJy will be about 1.3\% of the total flux density of 449 mJy given in Table~2. Such a loss does not affect the spectral analysis performed in Sect.~3.4.

However, the new radio images, showing nothing especially new in respect to the
archival ones, extend the observational data from two to the five different
frequencies ranging from 334 MHz to 4860 MHz. This is the necessary and sufficient
condition for performing the undertaken analyses.

\begin{figure*}[t]
\centering
\includegraphics[width=180mm,angle=0]{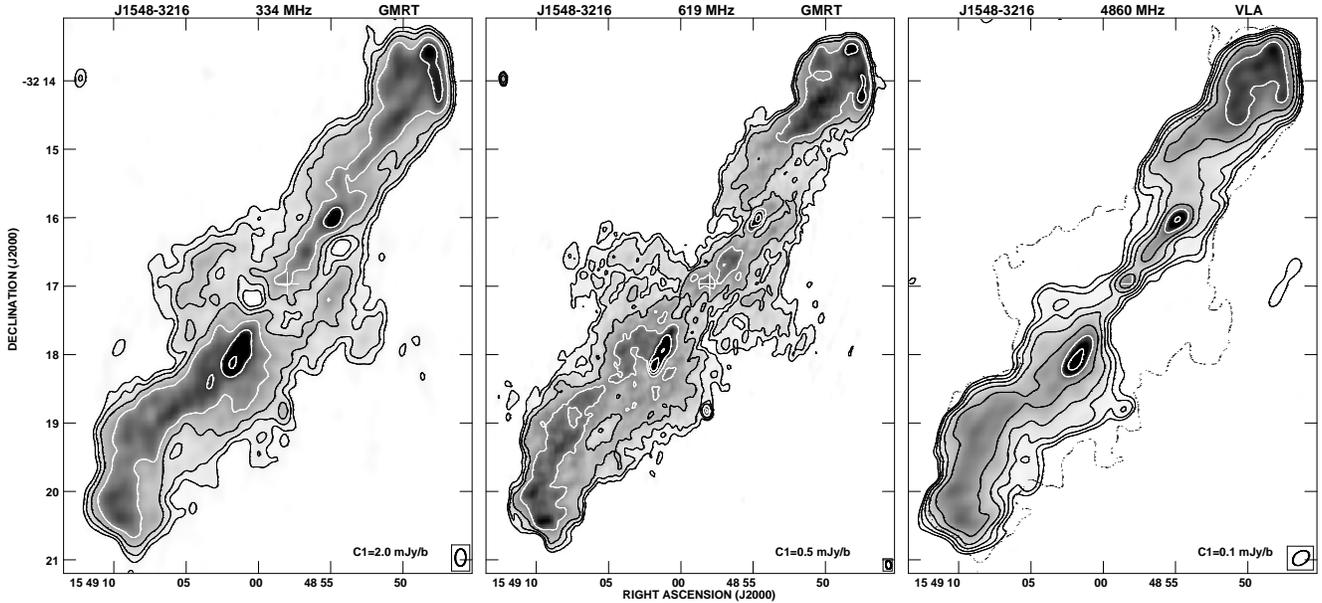}
\caption{Full-resolution GMRT 334 and 619 MHz, as well as VLA DnC-array  4860 MHz
images of the entire structure of the radio galaxy J1548$-$3216 (PKS 1545-321). The
first contour level, C1, is given in each image. The contour levels are (1, 2, 4,
8,...)$\times$ C1 mJy/beam.
In all the images, the restoring beam is indicated by an ellipse. The cross indicates
the position of the optical parent galaxy. The dashed outermost contour on the
VLA image encloses an area of the flare of the emission bridge within which a missing
4.86 GHz flux density is estimated in the text.}
\end{figure*}
  
\subsection{Extraction of the inner double structure}

To perform the spectral-ageing and dynamical analyses separately for the
outer and the inner structures, we have to extract emission of the inner lobes from
the underlying `background' radiation of the outer lobes. At the observing
frequencies of 334 and 619 MHz this is made by excluding of the visibility data taken
with baselines shorter than 2 k$\lambda$ and 3 k$\lambda$, respectively, while at
4860 MHz visibilities with spatial frequencies less than 2.5 k$\lambda$ are excluded. This effectively resolved out a large part of the underlying bridge's emission. Somewhat different approach was applied at 2495 MHz. Having the final ATCA image at
this frequency only but not the original UV data, we used the AIPS task {\sc IM2UV}
which allows a Fourier transformation of the image reconverting the data back to a UV
data file. Then a similar procedure, as described above, was applied to the
reconverted 2495-MHz UV data excluding the visibilities at baselines shorter than 2.5 k$\lambda$.

The resulting 334, 619, 2495, and 4860 MHz images of the inner double structure are shown in Fig.2. The corresponding archival ATCA 1384 MHz and VLA/BnA 4910 MHz images
are included for comparison. Besides the brightest parts with the leading heads of
the inner lobes, our new images confirm the presence of another weak emission region in
the inner NW lobe detected by SSS2003 in their 2495-MHz total intensity images of the
inner structure. Unfortunately, the dynamic range of our images is too low to detect
more of the connecting emission seen in their Fig.~6. We estimate that such a missing
flux is from about $10\%$ at 334 MHz to about $2\%$ at 4860 MHz. All these images,
except VLA/BnA, brought to a common scale using the {\sc AIPS} task {\sc HGEOM}
and convolved to the angular resolution of $16\arcsec\times 16\arcsec$, are used to
make a longitudinal section along the inner structure. Such `slices' at the five
observing frequencies are shown in Fig.3.  To avoid problems with any missing flux,
we restrict our spectral and dynamical analyses of the inner lobes to their brightest
regions indicated in Fig.~3.

\begin{figure*}[t]
\centering
\includegraphics[width=125mm,angle=0]{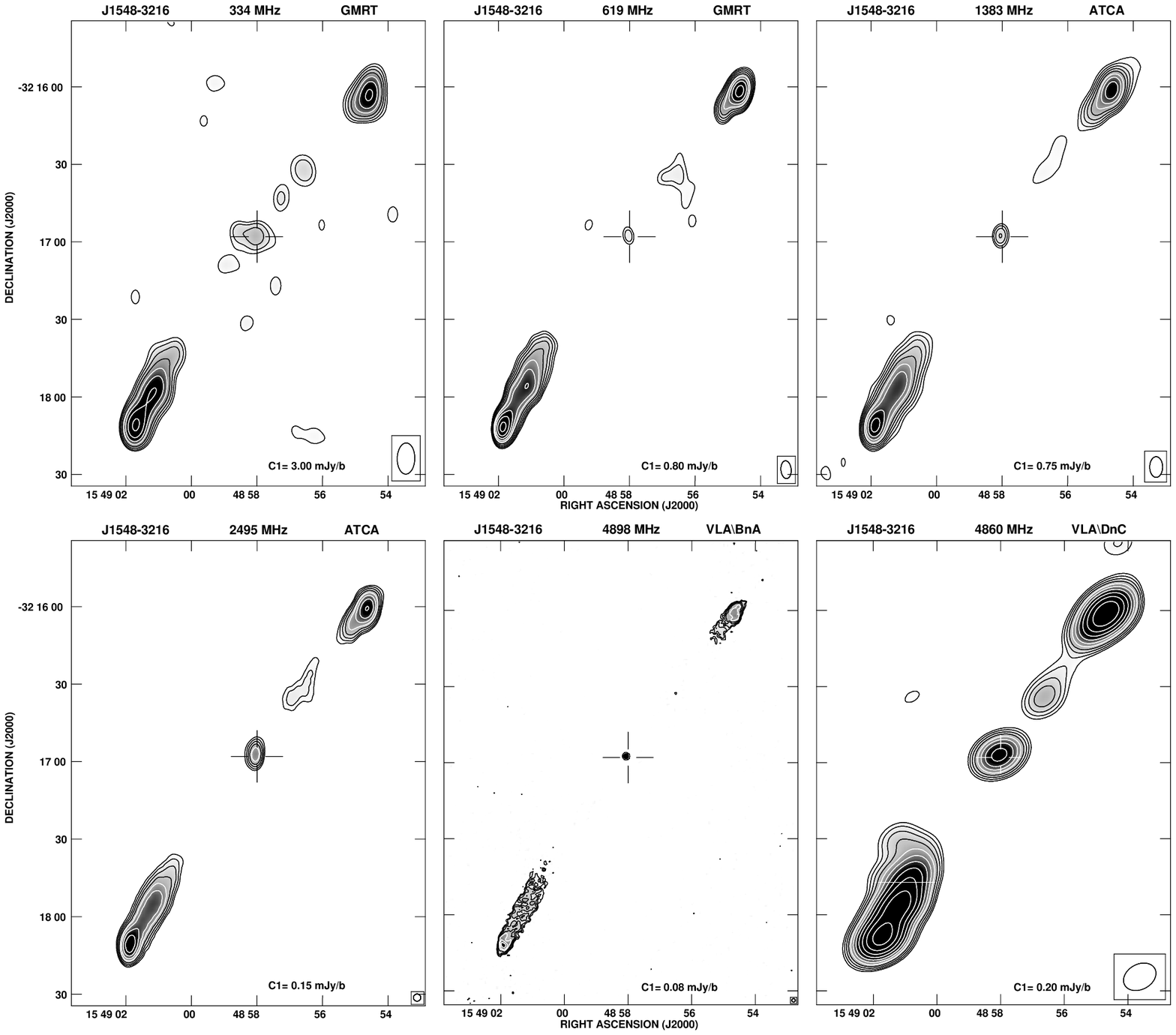}
\caption{GMRT, ATCA and VLA images of the inner double structure of J1548$-$3216. The
first contour level is given in each image. The contour levels are (1, 1.41, 2, 2.83,
4, 5.66,...)$\times$C1 mJy/beam. The restoring beam is indicated by an ellipse. The
cross marks the position of the optical galaxy.}
\end{figure*}

\begin{figure}[h]
\centering
\includegraphics[width=80mm,angle=0]{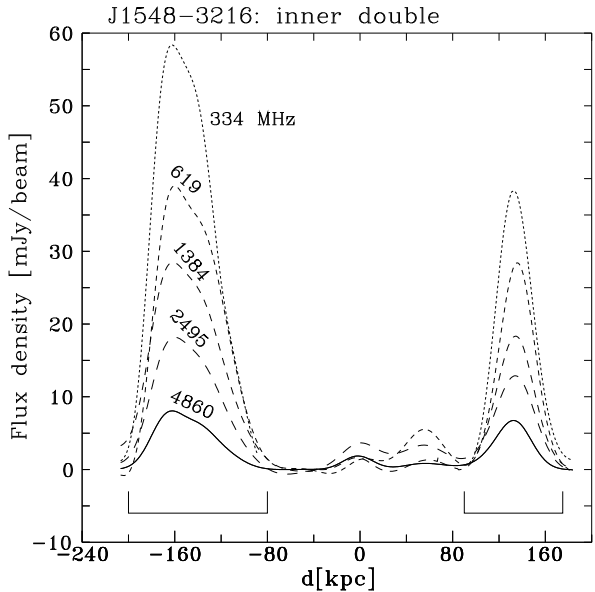}
\caption{Longitudinal section along the inner double structure. The horizontal brackets indicate regions of the structure subject to the spectral and dynamical analyses.}
\end{figure}

\subsection{Outer lobes cleaned from the inner structure}

To analyse physical properties of pure outer lobes of J1548$-$3216, the
inner double structure was subtracted from the images of the entire radio source
partly shown in Fig.~1. For this purpose, all of those images were also brought to a
common scale (a map size, cell size, coordinates of the map centre) and convolved to
the angular resolution of $16\arcsec\times 16\arcsec$. The images of the inner
structure  were blanked over regions outside the extracted inner lobes and then
subtracted from the convolved maps of the entire source using the AIPS task {\sc
COMB}. The net images of the outer lobes (rotated by $35\degr$) are shown in Fig.~4.
On the first of these images, a division of the radio structure into 18 strips, each
of them $28\arcsec$ wide, is shown. The first plotted contour on the ATCA+VLA image is
exceptionally high (about $8\times$ rms noise level) to clear it from spurious jagged
contours that appeared after the convolution of the original map with the beam of $16\arcsec\times 16\arcsec$. The integrated flux densities measured in the consecutive strips and plotted vs distance of the strip's centre from the core position (the strips' centres are separated by the angular distance of $28\arcsec/\cos 35\degr$) form a longitudinal section along the cleaned outer structure shown in Fig.~5a. The flares or spurs in the central region of the outer structure increase the total flux density in the strips S7, S8 and N11, N12 causing its peaks marked F1, F2, and F3 in Fig.\,5a. The brightness peaks of the leading heads of the new inner lobes lying at the
radio axis (indicated with the dotted vertical lines), almost coincide with the
positions of these strips. A spectral steepening and the spectral age within the
strips are analysed in Sect.~4.1.

\subsection{Radio spectra}

The integrated flux densities of the total source, as well as of its outer and inner
lobes, are given in Table 2. All columns are self-explanatory; outNW and outSE
indicate the NW and SE lobes of the outer double structure,  while innNW and innSE --
the NW and SE inner lobes, respectively. Because a spectral fit, especially with the SYNAGE, is very sensitive to a lack of low-frequency data (cf. MJS2009), the 160-MHz flux densities of the outer lobes are estimated by subtracting 300 mJy (assumed flux density of the inner double at this frequency based at a spectral index of about 0.6, cf. Sect.~4.2) from the total flux density measured with the Culgoora array, and dividing the net flux between the two lobes in a proportion similar to those observed at the higher frequencies.

Distributions of the low-frequency $\alpha^{334}_{1384}$ and high-frequency
$\alpha^{1384}_{4860}$ spectral index vs distance from the core measured along the
axis of the outer structure cleaned from the inner lobes are shown in Fig.\,5b. The
wavy ridge and its side flares do not show any peculiarity in the spectral index
distribution shown in Fig.\,5b. Both the low-frequency and the high-frequency indices
exhibit a systematic steepening from the heads of the outer lobes towards the centre.
The low-frequency spectral index rises from $\sim$0.5 to $\sim$0.9, while the
high-frequency one steepens from $\sim$0.8 to $\sim$2.0 at the evident depression of
emission at the centre of the bridge. Such a large continuous steepening of the
spectra suggested a systematic increase in the synchrotron age of relativistic
particles enclosed in the old cocoon, i.e. an increase from the lobes' head towards
their flaring ends.

We do not attempt to analyse a distribution of the synchrotron age in directions
transverse to the main axis of the source, hence spectral index distributions over
the entire area of the outer structure are beyond the scope of this paper.
Nevertheless our data show similar spectral features as those seen in the map of the
spectral index $\alpha^{2495}_{1384}$ in SSS2003 (their Fig.\,3), i.e. the steepest
spectra appear at eastern ends of the strips S7 and S8, and at western ends of the
strips N11 and N12. We cannot confirm a distinctly steeper spectrum along the
southwestern edge of the outSE lobe appearing in their map, but at least something
similar is not pronounced in the spectral index $\alpha^{1384}_{334}$ or
$\alpha^{1384}_{619}$.

\begin{figure*}[t]
\centering
\includegraphics[width=160mm,angle=0]{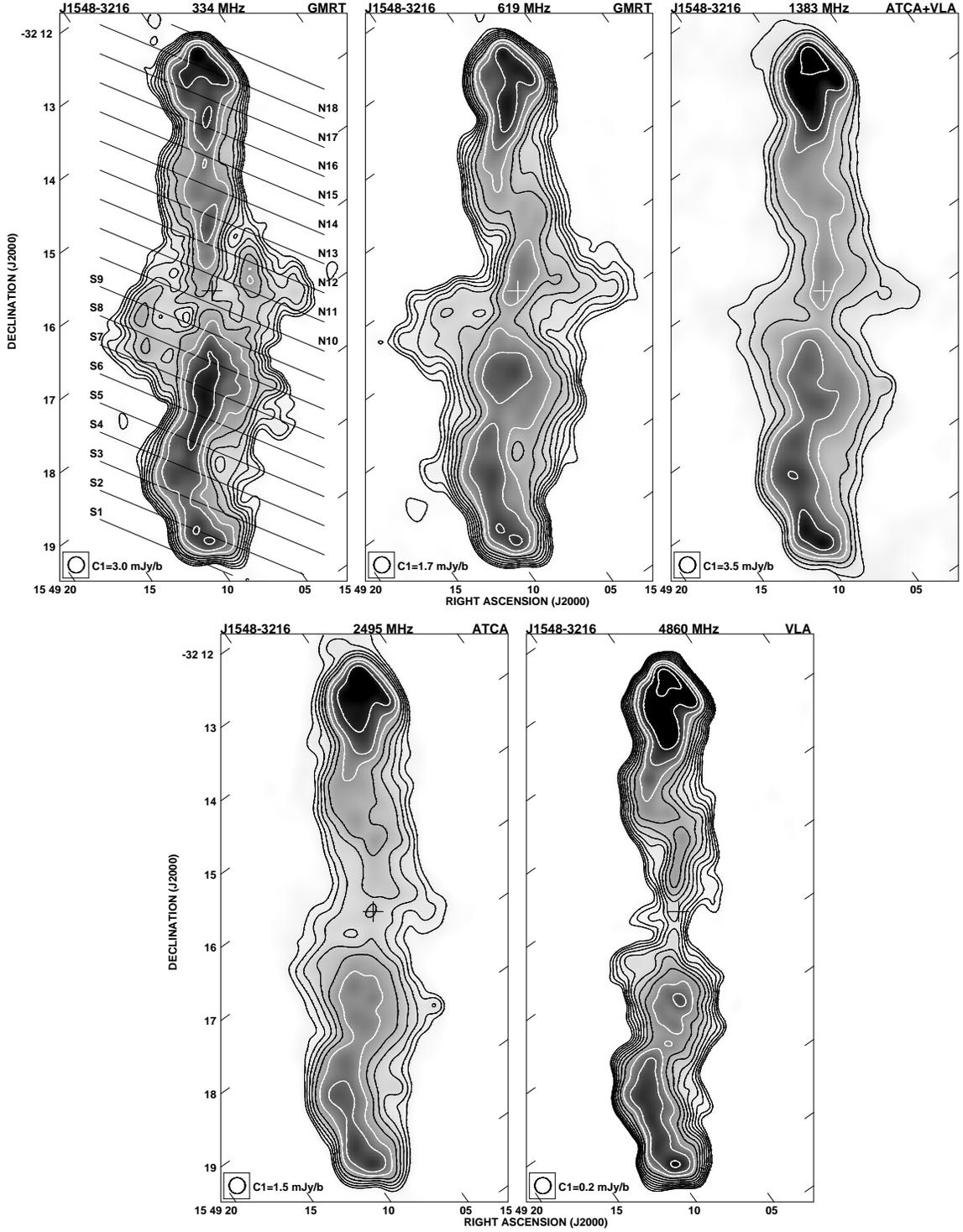}
\caption{Radio images of the outer lobes cleaned from the inner double structure and
convolved to the angular resolution of $16\arcsec\times 16\arcsec$. The first contour
level is given in each image. The contour levels are as in Fig. 2. The division into
18 $28\arcsec$-wide strips, used for the spectral steepening and spectral age
analysis, is shown in the first of these images.}
\end{figure*}

\begin{figure}[h]
\centering
\includegraphics[width=92mm,angle=0]{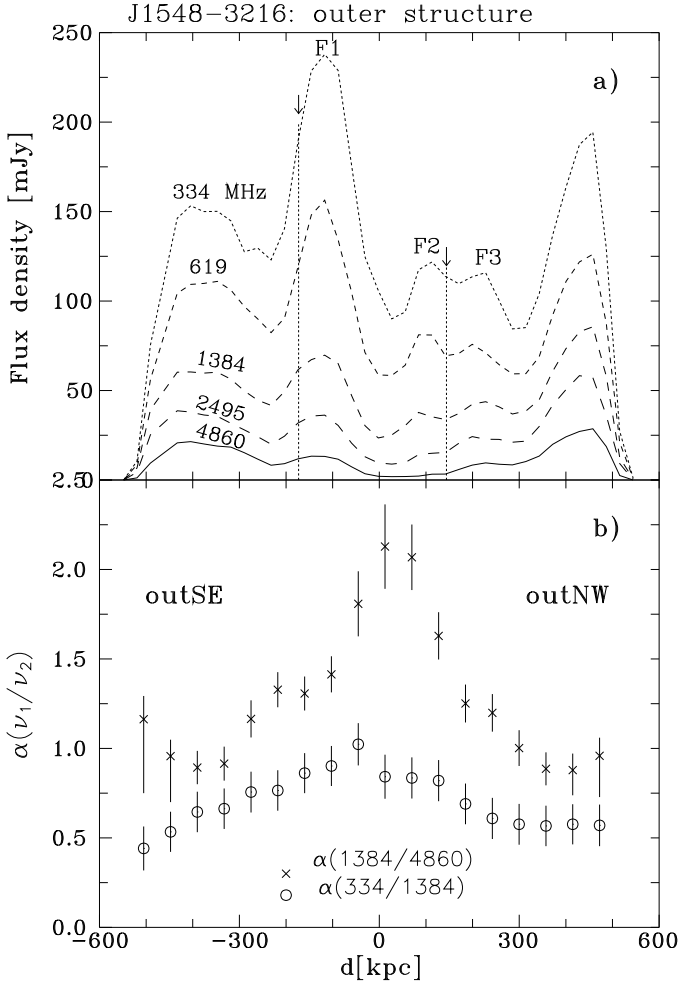}
\caption{{\bf a)} Integrated flux densities in the consecutive strips along the outer
double structure vs distance from the core. The vertical lines indicate positions of
brightness peaks in the inner lobes. {\bf b)} Low-frequency and high-frequency
spectral indices in these strips.}
\end{figure}

\begin{table*}[t]
\centering
\caption{Flux densities of the total structure (Total) and the outer (outNW and outSE)
and inner (innNW and innSE) lobes of J1548$-$3216.}
  \begin{tabular*}{123mm}{@{}rr@{}c@{}lr@{}c@{}lr@{}c@{}lr@{}c@{}rr@{}c@{}lc@{}c}
\hline
\hline
\multicolumn{1}{c}{Freq.} &\multicolumn{3}{c}{Total} &\multicolumn{3}{c}{outNW} &\multicolumn{3}{c}{outSE}
&\multicolumn{3}{c}{innNW} &\multicolumn{3}{c}{innSE}  & Ref.& \\
\multicolumn{1}{c}{[MHz]} &\multicolumn{3}{c}{[mJy]} &\multicolumn{3}{c}{[mJy]}&\multicolumn{3}{c}{[mJy]}
&\multicolumn{3}{c}{[mJy]} &\multicolumn{3}{c}{[mJy]}&  \\
\multicolumn{1}{c}{(1)} &\multicolumn{3}{c}{(2)} &\multicolumn{3}{c}{(3)}&\multicolumn{3}{c}{(4)}
&\multicolumn{3}{c}{(5)} &\multicolumn{3}{c}{(6)}& (7) \\
\hline
 160 & 8700 & $\pm$870& &(3900 & $\pm$&360)*&(4500 & $\pm$&480)* & & & & & & & (1) \\
 334 & 4926 & $\pm$740& & 2095 & $\pm$&314 & 2642 & $\pm$&396 & 58.0 & $\pm$&9.2 & 130.0 & $\pm$&19.7 & (5)\\
 619 & 3274 & $\pm$265& & 1383 & $\pm$&111 & 1758 & $\pm$&141 & 43.4 & $\pm$&4.6 &  87.6 & $\pm$&7.6  & (5)\\
 843 & 2519 & $\pm$200& & & & & & & & & & & & & & (2)\\
1384 & 1815 & $\pm$90 & &  820 & $\pm$&41  &  913 & $\pm$&46  & 26.5 & $\pm$&3.3 &  53.3 & $\pm$&4.0  & (4)\\
1400 & 1842 & $\pm$764  & & & & & & & & & & & & & & (3)\\
2495 & 1019 & $\pm$31&  &  479 & $\pm$&15  &  484 & $\pm$&15  & 17.5 & $\pm$&3.0 &  35.5 & $\pm$&3.2  & (4)\\
4860 &  449 & $\pm$25&  &  198 & $\pm$&20  &  217 & $\pm$&22  & 11.8 & $\pm$&1.6 &  19.7 & $\pm$&2.2  & (5)\\
4910 &      &      &    &      &      &    &      &      &    & 11.9 & $\pm$&0.9 &  23.1 & $\pm$&1.5  & (4)\\
\hline
\end{tabular*}

References: (1) Slee 1995; (2) Jones \& McAdam 1992; (3) NVSS (Condon et al. 1998);
(4) Flux densities measured on the archival ATCA and VLA images (Saripalli et al.
2003; Safouris et al. (2008); (5) this paper; ($^{*}$) estimated division of the
total 160 MHz flux density between the outer lobes, cf. the text.
\end{table*}   

\subsection{The radio core}

The J2000.0 position of the radio core determined from the high-resolution images is
R.A.: 15$^{\rm h}$ 48$^{\rm m}$ 58$^{\rm s}$.05 and Dec.: $-32\degr 16\arcmin
56\arcsec$.9, which is less than 1 arcsec away from centre of the parent galaxy
imaged with the Anglo-Australian Telescope (AAT) by S2008. The flux densities of the
core, measured on the images presented in this paper, are collected in Table 3. These
flux densities suggest a mildly inverted spectrum without a sign of time variability. 

\begin{table}[h]
\caption{Flux densities of the radio core}
\begin{tabular}{rcl}
\hline
\hline
Freq.  & S$_{\rm core}$ & Observing \\
  {[MHz]} & [mJy]  & date\\
\hline
 619 & 1.52 & 2008, Mar 8 \\
1384 & 2.17 & 2002, May 31 \\
2495 & 2.50 & 1999, Mar 7 \\
4860 & 2.53 & 2008, Jun 5 \\
4910 & 2.57 & 2002, Jul 30\\
\hline
\end{tabular}
\end{table}

\section{Spectral ageing analysis}

Remembering all the serious problems with both the principles and the practical
application of spectral-ageing calculations to physical conditions in radio sources
described in detail in MJS2009, the spectral age in different parts of the lobes,
i.e. the time elapsed since the radiating particles were last accelerated, is
determined using the classical theory that describes the time evolution of emission
spectrum of a single population of particles with an initial power-law energy
distribution (e.g. Myers \& Spangler, 1985; Carilli et al., 1991). The initial energy
distribution of the relativistic particles is a power-law function, 
$n(\gamma_{i})d\gamma_{i}=n_{0}\gamma_{i}^{-p}d\gamma_{i}$, of their initial Lorentz
factor, $\gamma_{i}$. The power $p$  corresponds to the initial (injection) spectral
index $\alpha_{\rm inj}$, which can be, in principle,  estimated from the
observational data until the synchrotron frequency of the minimum electron Lorentz
factor lies far outside the observable low-frequency spectrum. Fortunately, a
spectral turnover at low frequencies is not observed in the radio spectra of the extended FRII-type radio sources. On the other hand, the spectral break frequency above which the radio spectrum steepens from the injected power law, $\nu_{\rm br}$, is related to the spectral (synchrotron) age, $\tau_{\rm syn}$, and the magnetic field strength, $B$, through

\begin{equation}
\tau_{\rm syn}[{\rm Myr}]=50.3\frac{B^{1/2}}{B^{2}+B_{\rm iC}^{2}}[\nu_{\rm br}(1+z)]^{-1/2},
\end{equation}

\noindent
where $B_{\rm iC}$=0.318(1+$z)^{2}$ is the magnetic field strength equivalent to the
inverse-Compton microwave background radiation. $B$ and $B_{\rm iC}$ are expressed in
units of nT, while $\nu_{\rm br}$ is in GHz. Values of $\alpha_{\rm inj}$ and
$\nu_{\rm br}$ are found by the fit to the observed radio spectra using the SYNAGE
algorithm of Murgia (1996). 

\subsection{The outer structure}
\subsubsection{Determination of $\alpha_{\rm inj}$ and $\nu_{\rm br}$ values}

To determine the value of $\alpha_{\rm inj}$, we fit the CI, CIE, and JP
models to the flux densities of the entire outer lobes (given in columns 3 and 4 of
Table~2) treating $\alpha_{\rm inj}$ as a free parameter, and realizing that
decidedly the best fit to the data is achieved with the JP model. Fits of the JP
model of radiative losses to the flux densities of the outer SE and NW lobes are
shown in Fig.~6. The values of $\alpha_{\rm inj}$=0.583$^{+0.151}_{-0.070}$ and
$\alpha_{\rm inj}$=0.540$^{+0.096}_{-0.051}$ found by the fit correspond to the
$\alpha^{2495}_{1384}$ indices of $\sim$0.7 and $\sim$0.6 previously determined by
SSS2003 for the brightest regions at the SE and NW heads of the outer structure,
respectively. These fitted $\alpha_{\rm inj}$ indices are used to determine values of
$\nu_{\rm br}$ in the 18 parallel strips covering the entire outer structure of the radio source. The JP models of the spectra within these 18 strips are collected in Fig.~7.  A distance of the strip's centre from the core, the resulting value of $\nu_{\rm br}$, and the relevant value of $\chi^{2}_{\rm red.}$ giving a goodness of the fit in each of 18 strips, are given in columns 2, 3, and 4 of Table 4, respectively.

\begin{figure*}[]
\centering
\includegraphics[width=150mm,angle=0]{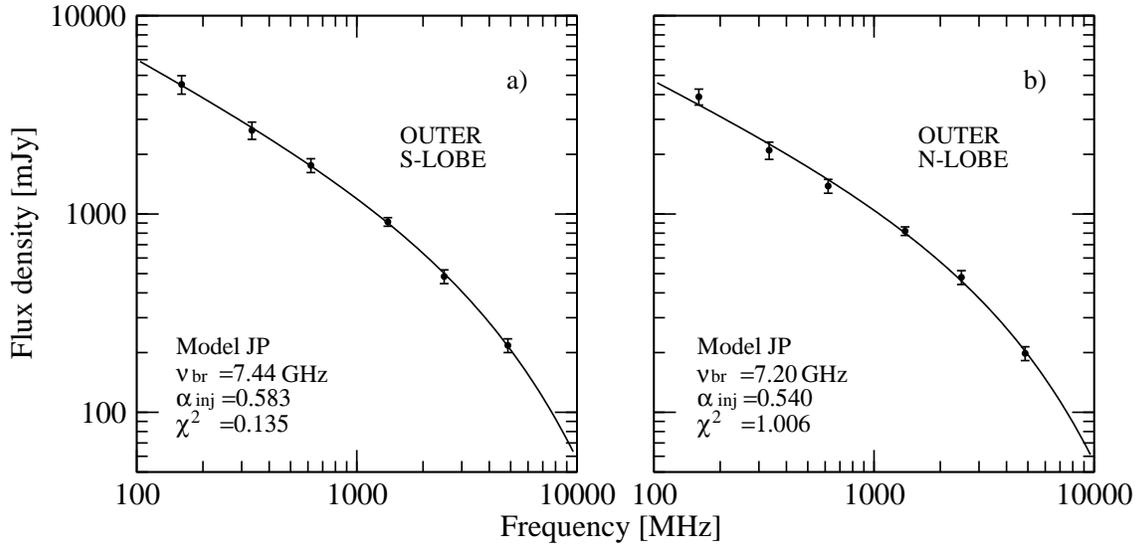}
\caption{Spectra of the outer lobes fitted with the JP model, as described in the text.}
\end{figure*}

\subsubsection{Determination of magnetic field strength values and the spectral ages}

In consistency with the approach applied in our previous spectral-ageing analyses of
giant radio galaxies (Jamrozy et al. 2005, 2008; Konar et al. 2006), the magnetic
field in Eq.\,(1) is identified with an `equipartition field', $B_{\rm eqv}$, which
provides equipartition between the total energy densities of the relativistic
particles and the magnetic field $(u_{\rm e}\approx u_{\rm B}$). The required values
of $B_{\rm eqv}$ are computed with Miley's (1980) prescription for the general formula

\begin{equation}
B_{\rm eqv}\propto (1+k)\left(\frac{L}{V}\right)^{2/7},
\end{equation}

\noindent
where $k$ is the ratio of the energy content of relativistic protons to that of
electrons (adopted as $k$=1), $L$ is the luminosity of a given strip calculated by
integration of its spectrum from a frequency equivalent to a minimum Lorentz factor,
$\gamma_{\rm min}\sim 1$ for the relativistic electrons to the upper limit of 100
GHz, and $V$ is the volume corresponding to that slice. The derived values of
$B_{\rm eqv}$ and the resulting spectral ages, $\tau_{\rm syn}$, are given in columns
5 and 6 of Table 4, respectively.

\begin{figure*}[]
\centering
\includegraphics[width=140mm,angle=-90]{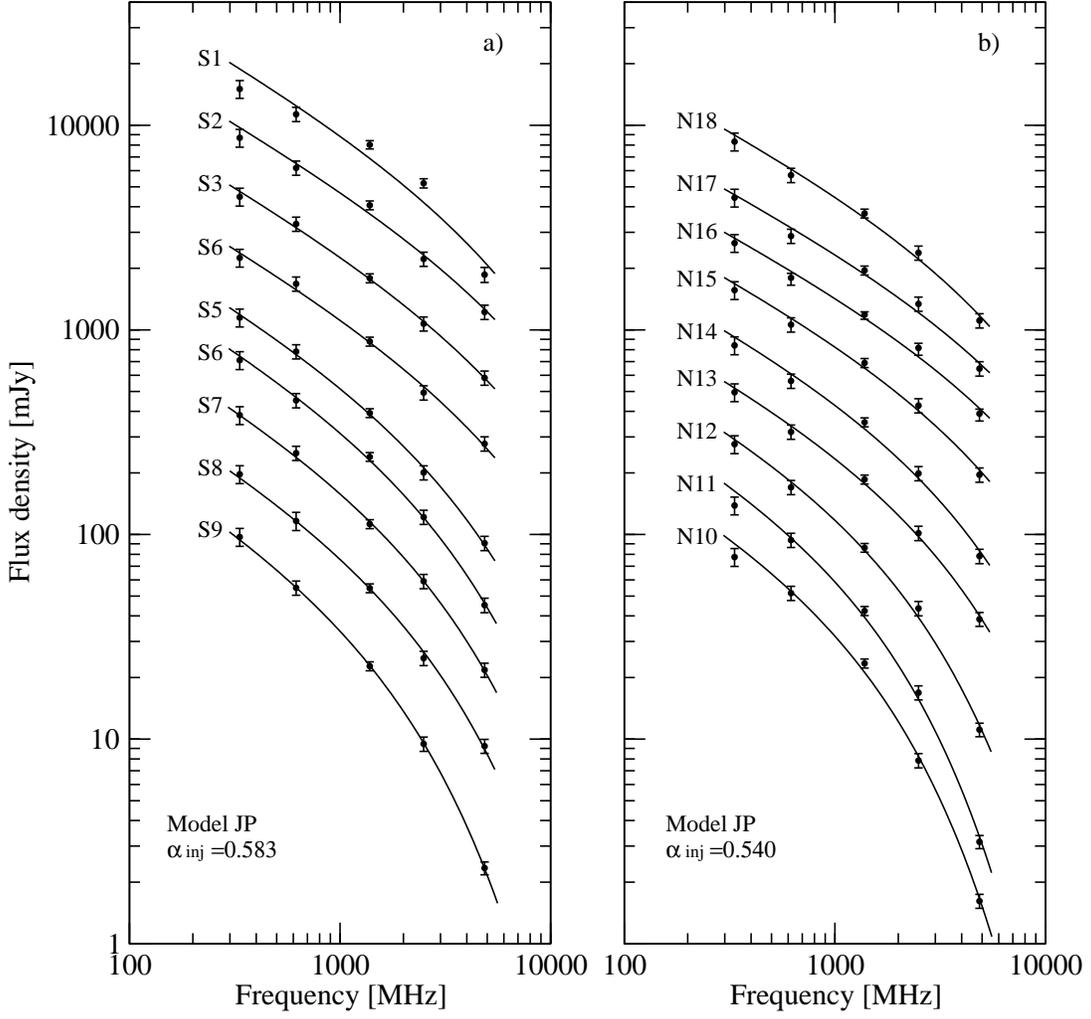}
\caption{Spectra of the slices S1 -- S9 in the outSE lobe and N10 --N18 in the outNW
lobe, fitted with the JP model. The spectra of particular strips are arbitrarily
shifted. Reduced values of $\chi^{2}$ and resulting values of the spectral break,
$\nu_{\rm br}$, are given in Table 4.}
\end{figure*}

\begin{table*}[t]
\centering
\caption{Break frequency, equipartition magnetic field strength, and spectral age of
emitting particles in consecutive strips through the outer lobes of J1548$-$3216
(cf. Fig. 4).}
\begin{tabular*}{115mm}{lcccccccc}
\hline
\hline
Strip & Dist.\,from & $\nu_{\rm br}$ && $\chi^{2}_{\rm red}$ && B$_{\rm eqv}$ && $\tau_{\rm syn}$ \\
    & core\,[kpc]   & [GHz] & & & & [nT] & & [Myr] \\
(1) &  (2)  & (3) & & (4) & & (5) & & (6) \\ 
\hline
    &        & outSE-lobe  &  &  &  &  $\alpha_{\rm inj}$=0.583 \\
S1  & $-$505 & 13.43$^{+3.24}_{-10.0}$  && 8.34 && 0.293$\pm$0.020 && 29.6$^{+11.2}_{-3.6}$ \\
S2  & $-$447 & 18.23$^{+7.80}_{-13.7}$  && 1.93 && 0.291$\pm$0.015 && 25.5$^{ +9.7}_{-5.5}$ \\
S3  & $-$390 & 15.79$^{+20.2}_{-8.41}$  && 0.26 && 0.269$\pm$0.014 && 27.7$^{ +7.3}_{-17.4}$\\
S4  & $-$332 & 13.37$^{+29.8}_{-5.36}$  && 0.65 && 0.260$\pm$0.014 && 30.3$^{ +5.9}_{-25.0}$\\
S5  & $-$275 &$\;\:6.93^{+4.76}_{-1.91}$&& 0.33 && 0.257$\pm$0.014 && 42.1$^{ +5.7}_{-14.0}$\\
S6  & $-$217 &$\;\:5.64^{+1.05}_{-1.58}$&& 0.35 && 0.242$\pm$0.012 && 46.9$^{ +6.3}_{-4.2}$ \\
S7  & $-$160 &$\;\:5.15^{+4.05}_{-0.88}$&& 0.21 && 0.238$\pm$0.011 && 49.1$^{ +4.0}_{-18.2}$\\
S8  & $-$102 &$\;\:4.42^{+4.29}_{-0.58}$&& 0.40 && 0.240$\pm$0.012 && 53.0$^{ +3.3}_{-24.4}$\\
S9  & $\;\:-$45 &$\;\:2.87^{+0.88}_{-0.31}$&&0.05&& 0.241$\pm$0.013&& 65.7$^{ +3.4}_{-9.6}$ \\
    & \\
    &        & outNW-lobe  &  &  &  &  $\alpha_{\rm inj}$=0.540 \\
N10 & $\;\:$+12 &$\;\:2.27^{+0.10}_{-0.46}$&&2.24&& 0.240$\pm$0.013&& 73.9$^{ +7.1}_{-1.7}$ \\ 
N11 & $\;\:$+70 &$\;\:2.38^{+0.10}_{-0.52}$&&2.19&& 0.240$\pm$0.012&& 72.2$^{ +7.5}_{-1.6}$ \\
N12 & +127   &$\;\:3.46^{+0.25}_{-0.79}$&& 1.04 && 0.238$\pm$0.012 && 59.9$^{ +6.5}_{-2.1}$ \\
N13 & +185   &$\;\:5.99^{+0.87}_{-1.87}$&& 0.77 && 0.256$\pm$0.014 && 45.3$^{ +6.9}_{-3.2}$ \\
N14 & +242   &$\;\:7.21^{+1.05}_{-2.98}$&& 1.72 && 0.264$\pm$0.015 && 41.1$^{ +8.3}_{-3.0}$ \\
N15 & +300   & 11.56$^{+2.94}_{-5.77}$  && 1.25 && 0.293$\pm$0.017 && 31.9$^{ +8.1}_{-4.1}$ \\
N16 & +357   & 17.32$^{+7.26}_{-10.2}$  && 1.28 && 0.309$\pm$0.018 && 25.7$^{ +7.9}_{-5.6}$ \\
N17 & +415   & 18.29$^{+8.42}_{-11.1}$  && 1.42 && 0.328$\pm$0.018 && 24.6$^{ +7.9}_{-6.0}$ \\
N18 & +472   & 13.17$^{+4.00}_{-7.00}$  && 1.19 && 0.330$\pm$0.021 && 28.9$^{ +8.2}_{-4.8}$ \\ 
\hline
\end{tabular*}
\end{table*}

The distribution of this spectral age vs distance from the core measured along the
axis of the outer structure is shown in Fig.\,8.

\subsection{The inner structure}

The spectrum of each of the two inner lobes, i.e. the flux densities given in columns
5 and 6 of Table 2, is fitted with the CI model. The fits (shown in Fig.\,9) suggest
a similar initial slope of the spectrum of both the lobes of $\sim 0.6\pm 0.1$ and
the spectral break of $161^{\rm dex(11)}_{-156}$ GHz and $44^{\rm dex(11)}_{-42}$ GHz
for the NW and SE lobes, respectively. (The formal $\pm 1\sigma$ errors are enormous
due to the practically straight spectra.) The volume of the lobes is calculated assuming their cylindrical geometry with a minimum angular size $24\arcsec\times 7\arcsec$ (height $\times$ base diameter) for the NW lobe and $42\arcsec\times 8\arcsec$ for the SE lobe where these dimensions are measured in the VLA image of S2008 (their Fig.~5). In this case, the equipartition magnetic field strength, calculated with the prescription of Miley (1980), is $0.63\pm 0.16$ nT and $0.65\pm 14$ nT, respectively.
Using these values, a `mean' spectral age of the radiating particles in the lobes is $5.4^{+2.8}_{-4.9}$ Myr for innNW lobe and $10.1^{+5.0}_{-9.5}$ Myr for innSE lobe. However, adopting the full length of the lobes as the cylinder's height, i.e. $74.\arcsec 6$ and $89.\arcsec 5$ for innNW and innSE lobes, the magnetic field strengths reduce to $0.46\pm 0.10$ nT and $0.43\pm 0.10$ nT, while the ages increase to $7.0^{+3.0}_{-6.5}$ Myr and $13.9^{+6.1}_{-13.3}$ Myr, respectively. The resulting spectral ages of the inner lobes are discussed in Sect.~6.1.

\begin{figure}[]
\centering
\includegraphics[width=92mm, angle=0]{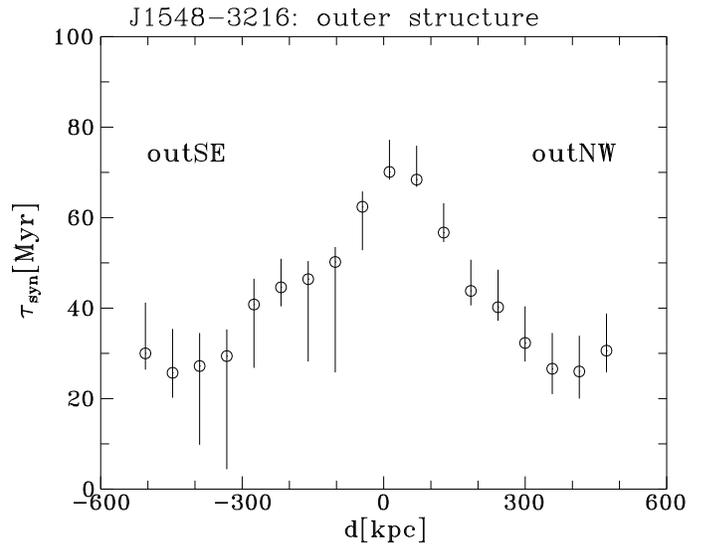}
\caption{Spectral age of relativistic particles in the outer structure cleaned from
the inner lobes plotted vs distance from the radio core.}
\end{figure}

\section{The dynamical age analysis}

This analysis is performed using the DYNAGE algorithm of Machalski et al. (2007)
which is based on the analytical model for the evolution of FR II type radio sources,
combining the pure dynamical model of Kaiser \& Alexander (1997) with the model for
expected radio emission from a source under the influence of the energy loss processes
published by Kaiser et al. (1997, known as the KDA model). With this algorithm we
derive the dynamical age of the lobes $t$, both the outer and the inner ones, the
{\sl effective} injection spectral index $\alpha_{\rm inj}$, which approximates the
initial electron continuum averaged over a very broad energy range and over the
present age of the source, the jet power $Q_{\rm jet}$, and the central density near
the radio core $\rho_{0}$, which determines the ambient density in which the jet
propagates. 

A detailed description of how to apply the above algorithm is published in MJS2009.
It is worth explaining here that determining of values of these four free
parameters of the model is possible by a fit to the observational parameters of a
source (or its lobes): its projected linear size $D$, the volume $V$, the radio power $P_{\nu}$ and the radio spectrum $\alpha_{\nu}$, which provides $(P_{\nu})_{\rm i}$
at a number of observing frequencies $i$=1, 2, 3,...  As in the KDA model, we assume
a cylindrical geometry of the lobes (cocoon), thus $V=\pi\,D^{3}/4\,R_{\rm T}^{2}$
where $R_{\rm T}$ is their axial ratio. The values of the few other free parameters of the model have to be assumed. These are the central core radius $a_{0}$, the exponent $\beta$ describing the ambient density profile in the simplified King's (1972) model $\rho(d)=\rho_{0}(d/a_{0})^{-\beta}$, the Lorentz factors determining the energy range of the relativistic particles used in integration of their initial power-law distribution $\gamma_{\rm i,min}$ and $\gamma_{\rm i,max}$, the adiabatic indices of the three `fluids' with individual energy densities: the jet material, $\Gamma_{\rm jet}$, the magnetic field, $\Gamma_{\rm B}$, and the ambient medium, $\Gamma_{\rm x}$ (cf. Kaiser et al. 1997). Since the emitting region consists of these three fluids, the model also takes the adiabatic index of the lobe (cocoon) into account as a whole, $\Gamma_{\rm c}$. The two other free parameters  we have to assume are $k^{\prime}$ -- the ratio of the energy density of thermal particles to that of the relativistic electrons -- and $\theta$, the orientation of the jet axis to the observer's line of sight. Following KDA, in the DYNAGE algorithm the assumed energy equipartition is expressed by the ratio of the energy densities of the magnetic field and of the particles, $\zeta=(1+p)/4=(1+\alpha_{\rm inj})/2$. The values adopted for the whole source are $a_{0}$=10 kpc, $\beta$=1.5, $\gamma_{\rm i,min}$=1, $\gamma_{\rm i,max}$=$10^{7}$, and $\theta$=90$\degr$. A decrease in $\theta$ to $\sim$70${\degr}$ (cf. S2008) would result in $\sim 6\%$ increase of $D$ and $\sim 7\%$ increase of $t$ (cf. Eq.\,(6) in Section 5.3). As we are interested in an age difference between the lobes rather than in their absolute age value, the latter one is less important. The values of $\Gamma_{\rm c}$, $\Gamma_{\rm B}$, $\Gamma_{\rm x}$, and $k^{\prime}$ assumed for
the outer and the inner lobes are given in next sections. The observational data
of these lobes, used in the DYNAGE fitting procedure, are given in Table~5. Most
columns are selfexplanatory, the entries in columns (6)--(9) give the ratios of the
size and luminosity of the given lobes.

Given the values of $\alpha_{\rm inj}$, $Q_{\rm jet}$, $\rho_{0}$, and $t$, several
other physical parameters of the source can be specified, e.g. the internal pressure
in the lobes $p_{\rm c}(t)$, their energy density $u_{\rm c}(t)$, a ratio of the
kinetic energy delivered by the jet to the energy radiated out,
$(Q_{\rm jet}\times t)/(u_{\rm c}\times V)$, and an average expansion speed of the
lobes, $D/(c\times t)$. The assumption of the energy equipartition condition allows
the magnetic field density $u_{\rm B}(t)$ and the field's strength $B(t)$ to be
estimated. The detailed expressions were given in our previous paper (cf. MJS2009). The
table with all notations for physical parameters used through the paper is given in
the Appendix.

However, the age and other physical parameters, fitted independently for either lobe
of a given double source, may be significantly different; i.e., any difference between
the fitted values of a parameter is found to be greater than the uncertainty of the
fits. This is a consequence of the usual asymmetries between the lobes in their
length and luminosity. The difference arises if the same density profile of the
ambient medium along the opposite lobes is assumed. The ratios between these
parameters of the lobes of J1548$-$3216, both the outer and the inner ones, are shown
in columns (6)--(9) in Table 5. On the other hand, we can expect that $Q_{\rm jet}$ and $\rho_{0}$ have the same values in the solutions for the opposite lobes, since they characterize an energy-emitting process in the central AGN. Also a large difference in age is rather unlikely. Therefore, following a similar ageing analysis in MJS2009, we consider the {\sl independent} solutions, as well as the two {\sl self-consistent} solutions for the opposite lobes, hereafter denoted as solutions A and B.

\begin{figure*}[]
\centering
\includegraphics[width=150mm, angle=0]{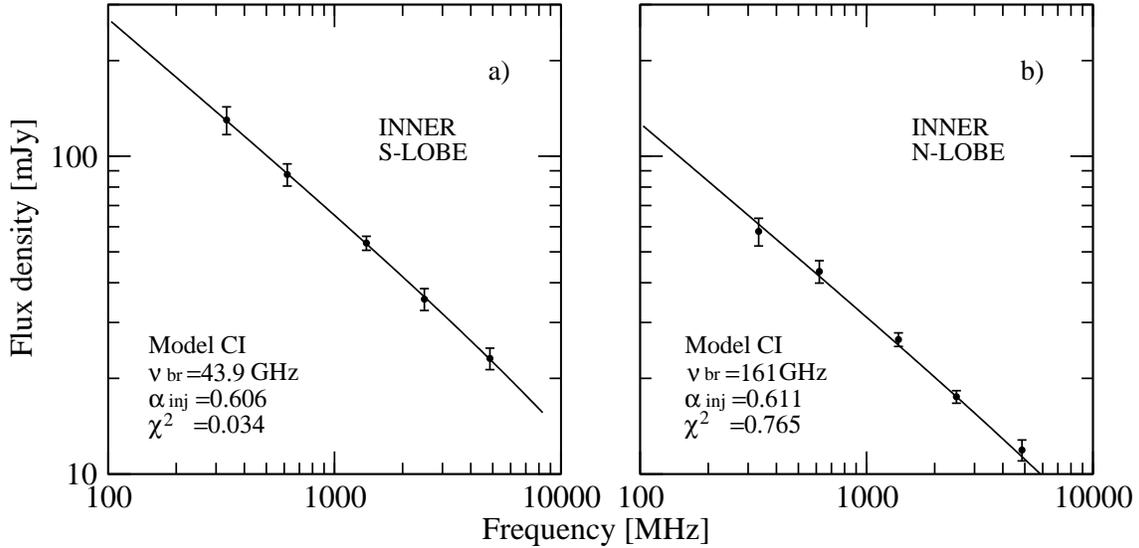}
\caption{Spectra of the inner lobes fitted with the CI model, as described in the text.}
\end{figure*}

\begin{table*}
\centering
\caption{Observational parameters of the outer and the inner lobes used to fit the
 dynamical model. The monochromatic powers are expressed in W\,Hz$^{-1}$sr$^{-1}$). }
\begin{tabular*}{137mm}{lcccccccc}
\hline
\hline
             &\multicolumn{4}{c}{$------$Lobes$------$} & SE  & NW  & outer  & inner\\
Parameter    & outSE  & out NW  & innSE  &  innNW & out/inn & out/inn & SE/NW & SE/NW\\
(1)          & (2)    & (3)     & (4)    &  (5)   &  (6)    &  (7)    &  (8)  &  (9)\\
\hline
$D$[kpc]       & 513    & 485    & 173    & 144   & 2.96  & 3.37 & 1.06  & 1.20 \\
$R_{\rm T}$    & 2.7    & 3.1    & 11.8   & 10.6 \\
log\,$P_{334}$ & 24.756 & 24.654 & 23.470 & 23.115 & 19.32 & 34.59 & 1.26 & 2.26\\
log\,$P_{619}$ & 24.585 & 24.481 & 23.300 & 22.991 & 19.27 & 30.90 & 1.27 & 2.04\\
log\,$P_{1384}$& 24.310 & 24.263 & 23.085 & 22.780 & 16.79 & 30.41 & 1.11 & 2.02\\
log\,$P_{4860}$& 23.716 & 23.668 & 22.705 & 22.432 & 11.46 & 17.22 & 1.12 & 1.68\\
\hline
\end{tabular*}
\end{table*}

\subsection{Independent solutions for the individual outer lobes}

The images in Fig.\,4 suggest that the diffused outer lobes may comprise a fraction
of thermal particles, thus the lobes, as a whole, may have a non-relativistic
equation of state. Therefore, we assume $\Gamma_{\rm c}$=$\Gamma_{\rm B}$=$\Gamma_{\rm x}$=5/3 and $k^{\prime}$=10. The latter value is less than $\sim$(25--140)
calculated by S2008 with the assumption that the hot spot pressure, $p_{\rm hs}$,
equals a minimum pressure $p_{\rm min}$, though they expected that a true hot spot
pressure should be higher, i.e. $p_{\rm hs}\geq p_{\rm min}$. Indeed, when studying
the large-scale X-ray environment of selected FRII radio sources, Belsole et al. (2007)
find that the internal pressure in their lobes, $p_{\rm c}$, is $\sim$(1--5) times
higher that the minimum (equipartition) pressure. Considering that the ratio
$p_{\rm hs}/p_{\rm c}$ in the DYNAGE algorithm varies from $\sim$4 to $\sim$20
(depending on the value of $R_{\rm T}$), the assumed value $k^{\prime}$ is justified.
 
The model solutions, i.e. the parameter values resulting from the independent fits,
are listed in columns 2 and 3 of Table~6.

\subsection{Self-consistent solutions for the outer lobes}

The differences between the values of the model parameters for the opposite lobes
found in the independent solutions come from different environmental conditions
(and/or different magnetic fields) on both sides of the core region. For this reason,
in the first kind self-consistent solution (solution A), we averaged the values of
$Q_{\rm jet,out}$ and $\rho_{0,{\rm out}}$ found for the opposite outer lobes (given
in columns 2 and 3 of Table~6), and now treat them as the fixed free parameters of
the model, $\langle Q_{\rm jet,out}\rangle$=$1.115\times 10^{38}$ W and
$\langle\rho_{0{\rm out}}\rangle$=$5.445\times 10^{-22}$ kg\,m$^{-3}$, respectively.
Given these values, we can determine another value of $\beta$ for each of the two
lobes, hereafter denoted as $\beta_{\rm sc.A}$. To do that (following MJS2009), we equalize values of the ambient density at the head of the outer lobes resulting from the independent solution and the self-consistent solution A:
 
\[\rho_{\rm a,out}=\rho_{0,{\rm out}}(D/a_{0})^{-\beta}\approx\langle\rho_{0,{\rm out}}\rangle 
(D/a_{0})^{-\beta_{sc.A}},\hspace{10mm}{\rm hence}\]

\begin{equation}
\beta_{sc.A}\approx\frac{\log(\langle\rho_{0,{\rm out}}\rangle/\rho_{\rm a,out})}{\log(D/a_{0})}.
\end{equation}

\noindent
Given a value of $\beta_{sc.A}$, the expected age of a lobe in the frame of the
self-consistent solution is

\begin{equation}
t_{\rm out}\approx\left(\frac{D}{c_{1}}\right)^{(5-\beta_{sc.A})/3}
\left(\frac{\langle\rho_{0,{\rm out}}\rangle a_{0}^{\beta_{sc.A}}}{\langle Q_{\rm jet,out}\rangle}\right)^{1/3},
\end{equation}

\noindent
where $c_{1}$ is a dimensionless constant dependent on the values of $\beta$,
$\Gamma_{\rm j}$, $\Gamma_{\rm x}$, and $\Gamma_{\rm c}$, given by Eq.\,(25) in Kaiser
\& Alexander (1997). The model parameter values resulting from the self-consistent
solution A are listed in columns 4 and 5 of Table 6.

The data in Table 6 show that the difference between the lobes' ages inferred from
the solution A is greater than the found in the independent solution. This is not
what we would expect for the actual ages of the opposite lobes\footnote{In MJS2009
we showed that for some sources the solution A can diminish the age difference. This
usually happens if the shorter lobe is brighter than the larger one}. Another
alternative, self-consistent solution is plausible in which these ages are very
similar (especially if we suspect that an orientation of the jets' axis in giant
radio galaxies is close to $\theta\approx 90\degr$), and any differences between the
linear extent and luminosity of the lobes come from an inhomogeneity either in
density distribution of the ambient gaseous environment or in magnetic field. Since
significant differences between the jet power and the radio core parameters in the
opposite directions along the jets' axis are not plausible, in the self-consistent
solution B we assume the same values of $\langle Q_{\rm jet,out}\rangle$ and
$\langle\rho_{0,{\rm out}}\rangle$ for both outer lobes (as in solution A) and the
same age $\langle t_{\rm out}\rangle$=132 Myr, i.e. a mean of the ages determined in
the independent solution, and $a_{0}$=10 kpc. In such a scenario, a value of
$\beta_{sc.B}$ can be calculated from Eq.\,(4) substituting $\langle t_{\rm out}\rangle$ for $t_{\rm out}$ and $\beta_{sc.B}$ for $\beta_{sc.A}$. As a result,

\begin{equation}
\beta_{sc.B}(D)=
\end{equation}
\[\left\{3\log\langle t_{\rm out}\rangle +
\log\left(\frac{\langle Q_{\rm jet,out}\rangle}{\langle\rho_{0,{\rm out}}\rangle}\right)-5\log\left(\frac{D}{c_{1}}\right)\right\}/\log\left(\frac{a_{0}}{D/c_{1}}\right).\]

\noindent
This solution does not give an unequivocal result for the fit, because $c_{1}$ is a
rising function of $\beta$. As shown and discussed in MJS2009, the parameter
space within which opposite lobes of a significantly asymmetric source would have the
same age is usually large. Nevertheless, an example of a reliable solution B for
$\beta_{sc.B}$ (calculated with $c_{1}(\beta$=1.5)), the corresponding value of
$\alpha_{\rm inj}$, and other model parameters of the outer SE and NW lobes are shown
in columns 6 and 7 of Table 6, respectively, for a comparison. It is worth noting
that in this solution the product $\langle Q_{\rm jet,out}\rangle\times\langle t_{\rm out}\rangle$ does not provide a minimum of the jet kinetic energy that is assured in the independent solutions. However, such a minimum is always shallow (cf. Machalski et
al. 2007) and the above objection is not very important.

The data in Table~6 show that the values of the model free parameters do not differ
much in the three solutions considered. However, the identity of $Q_{\rm jet,out}$,
$\rho_{0,{\rm out}}$, and $t_{\rm out}$ in the opposite outer lobes postulated in the
solution B result in a greater difference between the values of $\rho_{\rm a,out}$ and
$p_{\rm c,out}$ than in the remaining solutions. In particular, it suggests more
than twice denser ambient environment around the head of the outer NW lobe than
around the head of the opposite SE lobe. These physical conditions seem to be
supported by the presence of a distinct pair of emission peaks along the bright rim
at the end of the NW lobe, while a similar emission is absent in the SE lobe.
Also the mean pressure, $p_{\rm c,out}$ in the NW lobe is about 30$\%$ higher than
in the SE lobe, and their ratio found in the solution B is the highest. We discuss
this point again in Sect.~6.2.

\begin{table*}
\centering
\caption{Fitted physical parameters of the outer lobes, with
brackets $\langle\;\rangle$ showing the values assumed within the given
solution, (cf. the text).}
\begin{tabular*}{130mm}{@{}l@{}crr@{}rr@{}r@{}rr@{}r@{}rr}
\hline
\hline
{Parameter} &
\multicolumn{3}{c}{Indepen. solution} &\multicolumn{4}{c}{Self-consist. solut.\,A} &\multicolumn{4}{c}{Self-consist. solut.\,B}\\
&&\multicolumn{1}{r}{outSE} &\multicolumn{1}{r}{outNW}&&\multicolumn{1}{r}{outSE} &\multicolumn{1}{r}{outNW}
& &\multicolumn{1}{r}{outSE} &\multicolumn{1}{r}{outNW} \\
(1)  && (2) & (3) && (4) & (5) && (6) & (7) \\
\hline
$\beta$, $(\beta_{sc.A})(\beta_{sc.B})$   && 1.50  & 1.50  && 1.494 & 1.507 && 1.586 & 1.402 \\
$\alpha_{\rm inj}$                        && 0.552 & 0.531 && 0.536 & 0.545 && 0.527 & 0.554 \\
$Q_{\rm jet,out}(\times 10^{38}$\,W)          && 1.211 & 1.019 &&$\langle$1.115$\rangle$ &$\langle$1.115$\rangle$ &&$\langle$1.115$\rangle$ & $\langle$1.115$\rangle$ \\
$\rho_{0,{\rm out}}(\times 10^{-23}$\,kg\,m$^{-3}$) && 5.585 & 5.297 && $\langle$5.445$\rangle$ &$\langle$5.445$\rangle$ &&$\langle$5.445$\rangle$ &$\langle$5.445$\rangle$ \\
$\rho_{\rm a,out}(\times 10^{-25}$\,kg\,m$^{-3}$) && 1.521 & 1.570 && 1.535 & 1.574 && 1.040 & 2.382 \\
$t_{\rm out}$(Myr)                           && 142   & 122   && 146   & 119   &&$\langle$ 132 $\rangle$   &$\langle$ 132 $\rangle$   \\
$p_{\rm c,out}(\times 10^{-13}$\,Pa)       && 1.472 & 1.459 && 1.393 & 1.552 && 1.291 & 1.687 \\
$B_{\rm out}$(nT)                            && 0.322 & 0.320 && 0.313 & 0.330 && 0.300 & 0.345 \\
$U_{\rm out}=u_{\rm c}\times V(\times 10^{53}$\,J)&& 1.884 & 1.200 && 1.784 & 1.280 && 1.624 & 1.399 \\
$D/(t_{\rm out}\times c)$                    && 0.0118& 0.0130&& 0.0115& 0.0133&& 0.0128& 0.0120 \\
\hline
\end{tabular*}
\end{table*}

\subsection{Independent solution for the inner lobes}

In the case of the inner structure, we assume that (i) the observed emission arises
from the narrow lobes (cocoon), not the restarted jets; (ii) the jets' and lobes'
material has a relativistic equation of state with $\Gamma_{\rm jet}=\Gamma_{\rm B}=\Gamma_{\rm c}=4/3$ with no thermal particles, thus $k^{\prime}$=0;  and (iii) the restarted jets propagate within rarefied and uniform (with $\beta$=0) medium of the relict outer cocoon formed by the old jets' material that passed through the jet terminal shock. Since the observed spectra of the inner lobes show no curvature below the frequency of 4.9 GHz, especially for the innNW lobe where the SYNAGE fit suggests $\nu_{\rm br}$ above 20 GHz (cf. Fig.\,9), the DYNAGE algorithm will not be able to find a unique solution for the dynamical age, i.e. to determine values of 
$Q_{\rm jet,inn}$, $\rho_{0,{\rm inn}}$, and $t_{\rm inn}$, even if a value of 
$\alpha_{\rm inj}$ is known. Its formal fit with the SYNAGE is 
0.606$^{+0.079}_{-0.096}$ and 0.611$^{+0.062}_{-0.086}$ for the innSE and innNW
lobes, respectively. Therefore for the purpose of DYNAGE calculations, we assume here
that a maximum value of $\alpha_{\rm inj}$ cannot exceed the values of 0.606 and
0.611, but can be as low as 0.510 and 0.525, respectively. Moreover, the width of the
inner cocoon can be larger than the lobes' widths determined from the images in
Fig.\,2, therefore we admit a twice larger width for these lobes corresponding to
$R_{\rm T}$=6.5 (instead of 11.8 and 10.6) supposing that the best age solution for
the innSE lobe lies within the model space parameters limited from one side by the
values $\alpha_{\rm inj}$=0.606 and $R_{\rm T}$=11.8, and from the other side by
$\alpha_{\rm inj}$=0.510 and $R_{\rm T}$=6.5. For the opposite innNW lobe, the
limiting pairs of the model parameters are $\alpha_{\rm inj}$=0.611,
$R_{\rm T}$=10.6 and $\alpha_{\rm inj}$=0.525, $R_{\rm T}$=6.5. 

The sets of solutions
resulting from the fit of the model's free parameters to the linear size and the radio
powers of the inner lobes (given in columns (4) and (5) of Table 5), are presented in
Fig.\,10. This diagram clearly shows that the spaces of model parameters for the opposite inner lobes do not overlap. Obviously the lobes' asymmetries in the luminosity and size are too large to allow a comparable age and jet power solution in the model. Moreover, a selection of adequate pair of $Q_{\rm jet,inn}$ and 
$t_{\rm inn}$ values is not possible until a value of $\rho_{0,{\rm inn}}$ is fixed by means of some additional constraint.  

Let us therefore consider the limiting values for the core density 
$\rho_{0,{\rm inn}}$ within the old outer lobes. On the one hand, the upper limit for
the cold gas density may therefore be provided by studies of the internal depolarization of radio emission produced by the extended lobes of FRII-type radio
galaxies. For example, Garrington \& Conway (1991) found that the product of the cold
gas number density and the lobes' magnetic field strength is on average
$n_{\rm g}\times B <0.5\times 10^{3}$ m$^{-3}$ nT. This, with the
$B_{\rm eqv}\approx 0.64$ nT determined for the inner lobes in Sect.~4.2, gives
roughly $\rho_{0,{\rm inn}}\approx m_{\rm p}n_{\rm g} < 1.3\times 10^{-24}$ 
kg\,m$^{-3}$. We note in this context that the above equipartition magnetic field strength is compatible with the typical values 0.3 nT$<B<$ 3 nT found by means of multiwavelength analysis of the non-thermal lobes' emission (e.g. Kataoka \& Stawarz 2005; Croston et al. 2005). However, the ambient gas density within the old cocoon of DDRGs is likely lower than that of the typical FRII-type sources with linear sizes comparable to those characteristic for the inner double structures. It can be supposed that $\rho_{0,{\rm inn}}<\rho_{\rm a,out}$, where 
$\rho_{\rm a,out}=\langle\rho_{0,{\rm out}}\rangle (D_{\rm out}/a_{0})^{-1.5}$. With 
$\langle\rho_{0,{\rm out}}\rangle\approx 5.4\times 10^{-23}$ kg\,m$^{-3}$ and
$D_{\rm out}\approx 500$ kpc, we have $\rho_{0,{\rm inn}}< 1.6\times 10^{-25}$ kg\,m$^{-3}$. This value, corresponding exactly to the baryon density of $10^{-4}$ cm$^{-3}$ ($10^{2}$m$^{-3}$) typical for the intracluster medium (ICM) of nearby ($z<0.2$) clusters of galaxies (Croston et al. 2008), can be considered as a very upper limit for
$\rho_{0,{\rm inn}}$.

On the other hand, the lower limit is provided by an amount of matter injected into
the outer lobes by the old jet, which is

\[\rho_{0,{\rm inn}}\approx m_{\rm p}n_{\rm p}>\frac{Q_{\rm jet,out}t_{\rm out}}{V_{\rm out}
\Gamma_{\rm jet}c^{2}(1+k)},\]

\noindent
where $m_{\rm p}$, $n_{\rm p}$, and $n_{\rm e}$ are proton mass, proton, and electron
number densities, respectively, and $k=n_{\rm e}/n_{\rm p}$. Allowing for mildly
relativistic value of $v_{\rm jet}$ on large scales and only slight electron
dominance in terms of number density within the outflow (Sikora \& Madejski, 2000;
Celotti \& Ghisellini, 2008), one can expect roughly $\Gamma_{\rm jet}(1+k)<10$, and
thus the lower limit for the gas density within the old cocoon
$\rho_{0,{\rm inn}}>10^{-30}$ kg\,m$^{-3}$. However, such a density is extremely low
and insufficient to explain the observed characteristics of the inner structure. 
Indeed, power of the jet propagating through the medium with $\rho_{0,{\rm inn}}\ga 10^{-30}$ kg\,m$^{-3}$ and $\beta=0$ (calculated from suitably transformed Eq.\,(4)) sufficient to expand a lobe with $R_{\rm T}\approx 11$ ($c_{1}\approx 4$) to the length $D\approx 150$ kpc with the speed v$_{\rm h}\la c$, i.e. during $t\ga 0.5$ Myr, is $\sim 5\times 10^{35}$ W. Therefore, in the framework of the KDA model, the power radiated out at the frequency of 1.4 GHz will only be about $\sim 1.4\times 10^{18}$ W\,Hz$^{-1}$sr$^{-1}$, i.e. almost five orders of magnitude lower than the observed one! Kaiser has (2000) already argued that some `additional material must have passed from the environment of the source through the bow shock surrounding the outer source structure into the cocoon'.

Perhaps  a better lower limit for the gas density results from the numerical
simulation of an intermittent jet by Clarke \& Burns (1991). They show that the
restarted jet must propagate through a thermalized material left by the old jet, and
this medium is about 40 times less dense than the initial ambient medium. In our case
$\rho_{0,{\rm inn}}>\rho_{0,{\rm out}}(D_{\rm inn}/a_{0})^{-1.5}/40$. Therefore we
estimate that $2.5\times 10^{-26}<\rho_{0,{\rm inn}}<1.6\times 10^{-25}$ kg\,m$^{-3}$.
Taking a mean of these extreme values, $\langle\rho_{0,{\rm inn}}\rangle=6.3\times 10^{-26}$ kg\,m$^{-3}$, we achieve further limits for the space of parameters in Fig.~10. The vertices of the solid-line-marked diamonds in Fig.\,10 indicate solutions of $t_{\rm inn}$ and $Q_{\rm jet,inn}$ for $\langle\rho_{0,{\rm inn}}\rangle=6.3\times 10^{-26}$ kg\,m$^{-3}$ and the four combinations between the values of $\alpha_{\rm inj}$ and $R_{\rm T}$: (i) 0.611 and 10.6; (ii) 0.611 and 6.5; (iii) 0.525 and 10.6, and (iv) 0.525 and 6.5 for the innNW lobe for instance. In particular, the marked areas indicate that the dynamical age of the inner lobes is within a range of 6--13 Myr, which is in accordance with the synchrotron age derived in Sect.~4.2.

The value of $\langle\rho_{0,{\rm inn}}\rangle$ adopted in our calculations is more
than two orders higher than its estimate in S2008 ($<5\times 10^{-28}$ kg\,m$^{-3}$)
and close to the ionized gas density in clusters of galaxies (cf. Croston et al.
2008). This seems to contradict the finding of SSS2003 that the investigated
radio galaxy is not in a cluster, and one could expect that the gas density inside the
old cocoon is much lower than the value $\langle\rho_{0,{\rm inn}}\rangle$ adopted in
the above calculations unless an efficient entrainment process occurs. However,
an order of magnitude decrease in this value, i.e. to about $6\times 10^{-27}$
kg\,m$^{-3}$, will result in a two-fold decrease in the estimated age of the inner
structure only, i.e. to about $\ga$3 Myr.

Nevertheless, there is a way to avoid any assumption about the environment's density.
It is not an easy task to detect and measure the high-frequency radio spectrum of the
inner lobes. The calculation of the flux density expected from the model with
$\alpha_{\rm inj}=0.56$, $Q_{\rm jet,inn}=1.26\times 10^{37}$ W, $\rho_{0,{\rm inn}}=
6.8\times 10^{-26}$ kg\,m$^{-3}$, $R_{\rm T}=11$, and $t=9$ Myr are 20.6 mJy, 11.0
mJy, 4.6 mJy, and 1.9 mJy at $\nu=$ 10.6, 30, 90, and 230 GHz, respectively.
Supplementing the observed spectrum with the above flux densities and fitting the CI
model of energy losses, we find $\nu_{\rm br}=110.4$ GHz, which with $B_{\rm eqv}=0.45$
nT gives $\tau_{\rm syn}=8.6$ Myr.

\begin{figure}[t]
\centering
\includegraphics[width=92mm,angle=0]{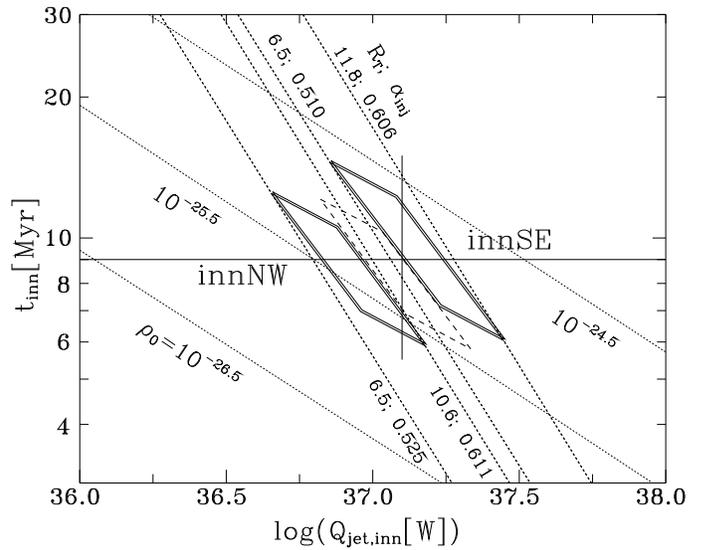}
\caption{Model solution for the inner lobes. The two pairs of dotted diagonal lines
for each of the lobes indicate the age vs jet-power relation corresponding to the
limiting values of the lobe's axial ratio, $R_{\rm T}$, and injection spectral index,
$\alpha_{\rm inj}$, marked in the diagram. The diamond-shaped solid-line areas show
credible dynamical age solutions for the two lobes. The dashed diamond indicates an
alternative solution for the innNW lobe providing an equalization of the jet powers for the opposite lobes (cf. the text). The solid vertical line indicates the value of
$Q_{\rm jet}$ adopted for the alternative age solution.}
\end{figure}

Similar to the outer lobes, the independent solutions for the inner lobes cannot
provide comparable jet powers for the opposite lobes. The diamond-shaped areas 
in Fig.\,10 are significantly separated along the abscissa. We therefore consider an
alternative solution of their age.

\subsection{Alternative solution}

An explanation of the difference in $Q_{\rm jet,inn}$ values by different external
density profiles along the opposite new jets is problematic if we assume a constant
density with $\beta$=0. The nearer opposite sides of the `diamonds'
correspond to the largest acceptable differences between the parameters $R_{\rm T}$
and $\alpha_{\rm inj}$. Even completely different values of these parameters for the
opposite inner lobes do not allow even a partial superposition of these diamonds
in the (log) plane $t_{\rm inn}$--$Q_{\rm jet,inn}$. However, as pointed out
and discussed in MJS2009, a plausible equalization of $Q_{\rm jet}$ values can be
achieved by changing the ratio between the energy densities of magnetic field and of
particles, $\zeta$. For example, the six-fold decrease of $\zeta$ in the innNW lobe
results in desired increase in $Q_{\rm jet,inn}$ in this lobe. This {\sl alternative}
solution is marked in Fig.\,10. We discuss this solution in Sect.~6.2.

Adopting the jet power, $\langle Q_{\rm jet,inn}\rangle=1.26\times 10^{37}$ W, which
intersects the diamond-shaped areas in the middle indicating the {\sl independent
solution} range for the innSE lobe and the {\sl alternative solution} for the innNW
lobe (Fig.\,10), and considering that
$\langle\rho_{0,{\rm inn}}\rangle= 6.3\times 10^{-26}$ kg\,m$^{-3}$, as well as
$\beta$=0, a probable age of the inner lobes estimate, $t_{\rm inn}$, is given by
Eq.\,(4) where the parameters denoted with `out' are replaced by those denoted with
`inn'. Given two different values of $c$1 corresponding to two extreme values of
$R_{\rm T}$ adopted here for either lobe, the age estimate is between 6.3 and 8.8 Myr,
and between 7.9 and 12.0 Myr for the innNW lobe and innSE lobe, respectively. 

A number of physical parameters fitted with the model for the inner lobes of
J1548$-$3216, with the ages as above, are provided in Table~7. Columns (2) and (3) give
their values for the innSE lobe with $R_{\rm T}$=11.8 and innNW lobe with
$R_{\rm T}$=10.6, respectively, while columns (4) and (5) give their values for these lobes but with $R_{\rm T}$=6.5. Although the fitted values of  $Q_{\rm jet,inn}$ and $\rho_{0{\rm inn}}$ differ a little from the values
$\langle Q_{\rm jet,inn}\rangle=1.26\times 10^{37}$ W
and $\langle\rho_{0,{\rm inn}}\rangle= 6.3\times 10^{-26}$ kg\,m$^{-3}$ used to
estimate the age from Eq.\,(4), their ratio $\rho_{0{\rm inn}}/Q_{\rm jet,inn}$
always equals  $\langle\rho_{0,{\rm inn}}\rangle / \langle Q_{\rm jet,inn}\rangle$.
This occurs because of the property of the DYNAGE algorithm allowing determination of
the values of both $\rho_{0}$ and $Q_{\rm jet}$ by the fit of the model free
parameters to the observed size and power of a given source (lobe). The last two lines in Table~7 give the expansion velocity of the jet's head resulting from the model and exhibit its deceleration with time. Since the length of the jet is

\begin{table*}[t]
\centering
\caption{Fitted physical parameters of the inner lobes for different values of the lobe's
axial ratio $R_{\rm T}$.}
\begin{tabular}{lrrccrr}
\hline
\hline
Parameter  &\multicolumn{2}{c}{Indepen. solution} & & &\multicolumn{2}{c}{Alternat. solution} \\  
            &\multicolumn{2}{c}{$---{\rm innSE}---$} & & &\multicolumn{2}{c}{$---{\rm innNW}---$} \\
(1)          & (2)      & (3)   & & & (4)       & (5)   \\
\hline
$R_{\rm T}$                              & 11.8  & 6.5   & & &10.6  & 6.5 \\
$\alpha_{\rm inj}$                       & 0.510 & 0.605 & & &0.525 & 0.610 \\
$Q_{\rm jet,inn}(\times 10^{37}$\,W)     & 1.53  & 1.21  & & &1.39  & 1.34 \\
$\rho_{0{\rm inn}}(\times 10^{-26}$\,kg\,m$^{-3}$)& 7.70 & 6.07  & & &7.17  & 6.13 \\
$t_{\rm inn}$(Myr)                       & 7.9   & 12.0  & & &6.3  & 8.8  \\
$p_{c,inn}(\times 10^{-14}$\,Pa)         & 2.91  & 3.34  & & &3.45  & 4.10 \\
$B_{\rm inn}$(nT)                        & 0.20  & 0.22  & & &0.22  & 0.24 \\
$U_{\rm inn}=u_{\rm c}\times V(\times 10^{50}$\,J) & 1.50  & 5.66  & & &1.27 & 3.44 \\
$D/(t\times c)$                          & 0.072 & 0.047 & & &0.075 & 0.053 \\
$v(r=a_{0})/c$                           & 0.286 & 0.189 & & &0.267 & 0.189 \\
$v(r=D_{\rm inn})/c$                     & 0.043 & 0.028 & & &0.045 & 0.032 \\
\hline
\end{tabular}
\end{table*}

\begin{equation}
r(t)=c_{1}\left(\frac{Q_{\rm jet}}{\rho_{0}a_{0}^{\beta}}\right)^{1/(5-\beta)}
t^{3/(5-\beta)},\hspace{10mm}{\rm implying}\hspace{1.5mm}{\rm that}
\end{equation}

\begin{equation}
v(t)=\frac{dr}{dt}=\frac{3\,c_{1}}{5-\beta}\left(\frac{Q_{\rm jet}}{\rho_{0}a_{0}^{\beta}}\right)
^{1/(5-\beta)}t^{(\beta -2)/(5-\beta)}, \hspace{10mm}{\rm or}
\end{equation}

\begin{equation}
v(r)=\frac{3\,c_{1}}{5-\beta}\left(\frac{Q_{\rm jet}}{\rho_{0}a_{0}^{\beta}}\right)^{1/3}
\left(\frac{r}{c_{1}}\right)^{(\beta -2)/3}.
\end{equation}

\noindent
The next to last line in Table 7 gives the expansion velocity at the assumed
radius of the radio core $a_{0}$, while the last one gives this velocity at the
actual length of the given inner lobe $D_{\rm inn}$.

\section{Discussion of the results}

In the sections below, we discuss the derived ages and physical parameters of the
source, both its outer and inner double structures, as well as those associated with
the original unperturbed ambient environment and the relict cocoon.

\subsection{Spectral age of the outer and the inner structures}

As expected, the synchrotron age of both the outer lobes, $\tau_{\rm syn}$, increases
with distance from their edges. It starts from $\sim$25 Myr and rises to about 65--75
Myr in the centre of the old cocoon. The lowest estimated age, far from the zero
value, can be related to the time that passed from the last acceleration of the
emitting particles at the region of the old jet interaction with the original
unperturbed IGM. However, in a number of similar spectral-ageing analyses (cf. Liu,
Pooley, Riley 1992; Jamrozy et al. 2005, 2008) a mean of the ratio between the
youngest and the oldest emitting particles is always about $0.40\pm 0.15$ and is weakly dependent on the size and/or the age of sources. It means that the youngest
synchrotron age is found usually between 5 Myr and 20 Myr for FRII-type sources with
$D\ga 100$ kpc and $t\ga 20$ Myr (cf. Machalski, Chy\.{z}y \& Jamrozy 2004). Similar
values ($\sim 18.7$ Myr and $\sim 20.3$ Myr) have been found in the detailed ageing
analysis of the DDRG J1453+3308 (Konar et al. 2006), for which the time that passed
since the jets stopped actively feeding the hot spots was estimated by Kaiser et al.
(2000) as about 1.1 Myr. We return to this point again in Sect.~6.3. 

The evidently curved age--distance tracks in Fig.\,8, especially for the outNW lobe,
support the expected deceleration of the jet head's advance speed and probable
backflow present in giant radio galaxies (e.g. Schoenmakers et al. 1998; Lara et al.
2000). Indeed, the advance speed of the jet's head given by Eq.\,(8),
$v(r)\equiv v_{\rm hs}$, and a backflow velocity, $v_{\rm bf}$, will form opposite
vectors. If a quotient of separation between the strips in Table~4 and a difference
between the derived spectral (synchrotron) ages in these strips,
$\Delta r/\Delta\tau_{\rm syn}=\langle v_{\rm syn}\rangle$, is more than $v(r)$
at a given distance from the core, a backflow is present. If 
$\langle v_{\rm bf}\rangle=\epsilon\langle v_{\rm hs}\rangle$, then
$\langle v_{\rm syn}\rangle=(1+\epsilon)\langle v_{\rm hs}\rangle$. For example, the
entries in Table~4 show that, in outSE lobe at $r=332$ kpc, $\epsilon\approx 2.2$ and
$v_{\rm bf}\approx 0.024\,c$, while closer to the core at $r=131$ kpc:
$\epsilon\approx 1.3$ and  $v_{\rm bf}\approx 0.017\,c$. In outNW lobe at
$r=300$ kpc, $\epsilon\approx 2.0$ and $v_{\rm bf}\approx 0.024\,c$, while at $r=98$
kpc: $\epsilon\approx 0.4$ and $v_{\rm bf}\approx 0.005\,c$.

A spectral age for the new inner lobes is quite problematic. Since this structure is
relatively smaller and much fainter than the outer structure, we were not able to cut
 it into strips and to check whether and how much the spectral index varies along the lobes. The spectral index map derived from the the ATCA+VLA and VLA images at 1384 and
2495 MHz, published by S2008 (their Fig.\,5), does not indicate any steepening that
would signal a spectral ageing. Though the flux densities measured in the entire inner
lobes (cf. columns (5) and (6) in Table~2) allowed the CI model fit to the data and an
estimate of spectral breaks in the observed spectra, the fit's uncertainty is so large
that the resulting `mean' spectral ages within the lobes are very uncertain as being
formally in a range from $\sim$0.1 Myr to $\sim$14 Myr. Taking an uncertainty of
the $R_{\rm T}$ value of the inner lobes into account (which determines their
volume, $V$, used to estimate $B_{\rm eqv}$), and that even the axial expansion of the
lobes with the speed of light would require the time of $\sim$0.5 Myr, we may adopt
that a spectral age of these lobes is about $9\pm 5$ Myr.
 
\subsection{Dynamical age of the outer and the inner structures}

Following  similar ageing analysis made by MJS2009 for a limited sample of ten giant
radio galaxies, we solved the dynamical age and other physical parameters of the two
outer lobes independently and explored the resulting formal differences between those
fitted parameters that should actually not differ in the opposite lobes, like the jet
power and the central core density. In this way we can perform a deeper search for
properties of the surrounding ambient medium and conditions of the jets' propagation
during the active phases of the jet production in the common central engine. In the
two kinds of proposed {\sl self-consistent} solutions, we require identical values
of $Q_{\rm jet}$ and $\rho_{0}$ for the opposite lobes both of the outer and of the
inner structures. However, in MJS2009 we showed that the fits with such an assumption
results in either decreasing or enlarging the age difference between the lobes,
depending on the actual asymmetries in the lobes' length and luminosity. The
{\sl self-consistent} solution A for the outer lobes of J1548$-$3216 enlarges the age
difference from $\sim$20 Myr in the {\sl independent} solution to $\sim$27 Myr (cf.
the entries in columns (2)--(5) in Table~6). As the above differences are too large to
be caused by a projection of the source on the sky, we consider the solution B in
which we demand the same age for both lobes. Taking a mean of the age values in the
independent solution, i.e. $\langle t_{\rm out}\rangle$=132 Myr, we find it as
satisfactory for the outer lobes. This is worth emphasizing that this age fully
agrees with its estimate in S2008, (30--200) Myr, though they derive it from another
consideration about the energy budget.

As expected, the {\sl self-consistent} solution B suggests different exponents
$\beta_{sc.B}$ in the power-law density profiles and different {\sl effective}
injection spectral index $\alpha_{\rm inj}$, where $\beta_{sc.B}$ value (according to
Eq.\,(6)) governs the lobe's length, while $\alpha_{\rm inj}$ value influences the
energy losses and the resulting brightness of a given lobe. Different fitted values
of $\beta_{sc.B}$ and $\alpha_{\rm inj}$ for the opposite outer lobes result in about
twice higher IGM density at the edge of the outer NW lobe as compared to that at the
SE lobe. As a result, the bright rim and the two `warm' spots in the outNW lobe may
indicate a higher pressure ratio between the lobe's head and the IGM, in other words,
this lobe may be more overpressured than the opposite outSE lobe.

The age for the inner structure is estimated by assuming a scenario in which
the new restarted jets propagate within rarefied and uniform ($\beta_{\rm inn}$=0)
medium of the old outer cocoon formed by the old jet material that passed through the
jet terminal shock and eventually partly mixed with the surrounding thermal gas (cf.
the discussion in Kaiser et al., 2000). The assumption of uniformity is justified by
the fact that the sound crossing time within the outer lobes is much longer than the dynamical timescales of evolution of the outer structure. In the case of the
inner structure, the age solution is formally undefined due to almost pure power-law
radio spectrum of the new lobes. On the one hand, such a lack of the spectrum
steepening can tell us that the structure is rather young; on the other hand, it
prevents determination of its quantitative value (cf. also SSS2003). In the framework
of the DYNAGE algorithm the inability to specify an explicit age solution means that
the space of the model's free parameters ($Q_{\rm jet}$, $\rho_{0}$, $t$,
$\alpha_{\rm inj}$) is very large (the space between two diagonal lines for a given
lobe in Fig.\,10). Within this space,  kinetic energy of the jet, 
$Q_{\rm jet}\times t$, is almost constant, so that the criterion of its minimum (a
crucial aspect of the algorithm) is not applicable. As shown in Sect.~5.3, the above
space can be reduced a little by fixing values of some of these parameters, e.g. the
value of $\rho_{0}$ (hence the diamond-shaped areas in Fig.\,10).

However, the straight, unaged spectra of the inner lobes are not just a single factor
precluding age determination. The other one is asymmetry both in their length and
brightness, which causes the spaces of parameters allowed for the opposite lobes
to not overlap each other. The assumption of uniform density of the pre-existing
cocoon with $\beta$=0 precludes a differentiation of this parameter for the lobes,
thus we pointed out another parameter that also rules the age solution, namely the
ratio between energy densities of the magnetic field and of the particles, $\zeta$.
The calculations indicate that a significant departure from the equipartition
conditions in one of the inner lobes (but not in the opposite one) enables a relative
equalization of the jet powers and attainment of comparable ages for both lobes.
Therefore, a combination of the {\it alternative} solution for the innNW lobe and the
{\it independent} one for the innSE lobe is required to attain comparable values of
the three parameters involved: $Q_{\rm jet,inn}$, $\rho_{0,{\rm inn}}$, and
$t_{\rm inn}$. Although this alternative age solution for the innNW lobe is only a
numerical result of the fitting algorithm, such a departure seems to be
supported by X-ray observations, which allow measurements of electron energies of
radio lobes and magnetic fields. In the lobes of the well-studied nearby radio
galaxies Centaurus\,B and Fornax\,A, the inferred electron energy densities exceed
those of magnetic fields by a factor of 5 -- 8 (cf. Tashiro et al. 1998, 2009; Isobe
et al. 2006).

Of course, the above combined age solution is odd, in which a significant departure
from equipartition conditions happens in only one lobe of the inner double structure
evolving in supposedly uniform medium, $\rho_{0,{\rm inn}}\approx$\,const.
Perhaps, these different conditions might arise when the magnetic field within the old
cocoon is strongly inhomogeneous giving rise to enhanced observable radiation in the
high-field regions (cf. Eilek et al. 1997; Kaiser 2005). An inhomogeneity of the field
may indicate a faint blob of emission detected at all the observed frequencies between
the radio core and the bright end of the innNW lobe (cf. Fig.\,2). Understanding an
expected criticism of the above speculations, we argue that the estimated dynamical
age of the inner lobes of the investigated radio galaxy, $t_{\rm inn}\approx 9\pm 4$
Myr, is plausible. This age accords with its spectral age estimate given in Sect.~5.3.

\subsection{Kinetic energy and internal pressure}

Kinetic energy delivered by the original jets to the old cocoon during the time of
$\sim$132 Myr is found to be $\sim$10$^{54}$ J, while the energy radiated out is
$\sim 1.3\times 10^{53}$ J (cf. entries in Table~6). The latter value is about twice
higher than estimated in S2008 using another approach to the energy budget. One
of the important results of our dynamical analysis is the power of the old and the new
jets. We find that the jet power during the previous phase of the nuclear activity was
$\sim 1.2\times 10^{38}$ W, while the power of the new jets (although we could only
determine a range of its value $\sim 5\times 10^{36}$ W to $\sim 3\times 10^{37}$ W;
cf. Fig.\,10)  seems to be almost one order less than the former one, but in a
good agreement with its range of [1.1, 4]$\times 10^{44}$ erg\,s$^{-1}$ if
$55\degr<\theta<78\degr$ estimated by S2008 (cf. their Fig.\,16). Such a high ratio
$Q_{\rm jet,out}/Q_{\rm jet,inn}$ impairs results of the ageing analysis of a few other
DDRGs published by Kaiser et al. (2000) in which they assumed the same $Q_{\rm jet}$
values during two successive phases of activity. The difference between the outer and
inner jet power found in our analysis may indicate that the spin of the central BH
cannot be the only factor determining the total power of the relativistic outflow, since
this spin is not expected to change significantly in a relatively short timescale,
which would be involved. However, we note that, in the original Blandford \& Znajek
(1977) model, the power of the outflow extracted from the ergosphere of the rotating BH
depends not only on the BH spin, but also on the magnetic field intensity within the
innermost parts of the accretion disk, which in turn should depend on the
accretion rate (or, possibly, on the accretion mode). Thus, the difference between the
power of the original and the restarted jet in the radio galaxy J1545$-$3216 may still
be reconciled with the modified spin paradigm (Sikora et al. 2007; Sikora 2009)
{\it if} the accretion disk undergoes significant structural changes between different
activity epochs in a single source.

The DYNAGE algorithm clearly shows that a demand of $Q_{\rm jet,\,out}\approx Q_{\rm jet,\,inn}$
would lead to an unlikely solution of the model for the inner structure with the lobes'
age $t_{\rm inn}\approx 0.8$ Myr, the old cocoon density
$\rho_{0,{\rm inn}}\approx 3.3\times 10^{-28}$ kg\,m$^{-3}$, and the average expansion
speed $D_{\rm inn}/t_{\rm inn}\approx 0.61\,c$!  One can conclude that until an
independent measure of $\rho_{0,{\rm inn}}$ is available (e.g. from internal
depolarization of radio emission and/or X-ray inverse-Compton emission from the region
of the inner lobes), a numerical algorithm like the DYNAGE will not be able to specify
more accurate values of $Q_{\rm jet,inn}$, $t_{\rm inn}$, $D_{\rm inn}/t_{\rm inn}$,
etc.

The internal (cocoon) pressures, $p_{\rm c}$, resemble the jets' disproportion. The
pressure in the old cocoon, derived from the fit, is about five times higher than the
pressure estimate within the new younger lobes (cf. the entries in Tables 5 and 7).
However, this is a result of the different equations of state assumed for the jet,
particle, and magnetic field `fluids': non-relativistic for the old cocoon and the
relativistic one for the inner double structure. If $\beta$=0, the cocoon pressure is

\[p_{\rm c}\propto \left(\frac{Q_{\rm jet}}{\rho_{0}}\right)^{2/5}\rho_{0}\,t^{-4/5}.\]

\noindent
For a lobe with given $D$ and $R_{\rm T}$ we find from Eq.\,(6) that $(Q_{\rm jet}/\rho_{0})\times t^{3}=$ const[m$^{5}$]. Thus

\[p_{\rm c}\propto \rho_{0}t^{-2}.\]

\noindent
Assuming the non-relativistic equation of state for a lobe at age $t$, the model
requires a higher jet power and predicts proportionally the higher core density needed
to fit the lobe's length and power, higher than the values of
$Q_{\rm jet}$ and $\rho_{0}$ fitted in the relativistic case. As a result, in the lobe
at age $t$, $p_{\rm c}\propto \rho_{0}$. The calculations show that even a large decrease in the lobe's age gives little increase of $p_{\rm c}$, so that the pressure
achieved in the non-relativistic case is not attainable.

\subsection{Magnetic fields}

With the standard assumption about the energy equipartition between the relativistic
particles and magnetic fields, the derived strength of uniform magnetic field in the
outer lobes, $B_{\rm eqv}$, slightly varies from $\sim$0.33 nT at the edges of the
lobes to $\sim$0.24 nT in the centre of the old cocoon. Similar values are derived
from the magnetic energy density, $u_{\rm B}$, via the dynamical considerations.
For the inner lobes, the field estimates are incompatible with the values derived
with the Miley (1980) formula during the spectral-ageing analysis, which are
almost twice higher than the corresponding values implied by the dynamical analysis.
Our calculations (in the framework of the KDA model) show that, to increase
strength of magnetic field in the inner lobes to its level derived from the spectral
analysis, it is necessary to assume the non-relativistic equations of state for the
jet's material and the magnetic field `fluids', as well as to admit some admixture of
thermal particles. Obviously such an assumption would level out the propagation
conditions during the initial and the restarted phases of activity, which is rather
unacceptable.  

\section{Conclusions}

The new low-frequency and high-frequency radio continuum maps of the double-double
radio galaxy J1548$-$3216 (PKS 1545$-$321) are used to complement its already
published maps at 22 and 13 cm (Saripalli, Subrahmanyan \& Shankar 2003; Safouris et
al. 2008) and to perform both the spectral-ageing and the dynamical analysis of this
remarkable giant DDRG, in which the newly restarted jets propagate through the bright
cocoon formed by a previous active phase of its AGN. The current activity is indicated
by the radio core.

The maps of the outer and the inner radio lobes at five observing frequencies from 334
to 4860 MHz allow determination of the spectral index distribution along the outer
lobes, as well as extension of the spectral index frequency range in the inner
lobes. The spectral index distribution in the outer lobes implies a distribution of
the spectral (synchrotron) age of emitting particles along the ridge of these lobes,
which increases from about 25 Myr at the edges of the extended old cocoon structure to
about 65--75 Myr in the vicinity of the core. A velocity equivalent to the quotient of
separation between the selected positions on the ridge and the difference in
synchrotron age between these positions, which is higher than the advance speed of the jet's head, indicates a significant backflow in the old cocoon. Moreover, its average speed evidently decreases from the cocoon's edges towards the core.

Using the DYNAGE algorithm (Machalski et al. 2007), we attempted to specify the
dynamical age of both the outer and the inner pairs of radio lobes. Because each pair
is characterized by a specific asymmetry in the lobes' length and brightness,
we searched for a self-consistent solution of the analytical model, Demanding the same
values for the jet power, central core density, and a comparable age for the opposite
lobes, we find that

(1) The age of the outer and the inner lobes is 132$\pm$28 Myr and $\sim$9$\pm$4 Myr,
respectively. However the derived age of the inner lobes is quite problematic because
the almost pure power-law spectrum of its lobes prevents a unique age solution
without any independent knowledge of either the restarted jet power or the ambient
density within the old cocoon through which it propagates. This problem is solved with
the assumption of limiting values for the latter parameter discussed in detail in
Sect.~5.3.

(2) If the above assumption is correct, the jet powers during the initial and the
restarted phase of activity are not similar. The restarted jet is one order less
powerful than the original one. We think that this can support a hypothesis that the
jet power is much more dependent (if not only) on the intermittent jet activity, than on the spin of a black hole. This intermittent activity can be connected with a stochastic transition between two accretion modes -- the standard one where the angular momentum is transmitted outwards by viscous torques within a disk and -- the `magnetic' one, in which large-scale poloidal fields are developed (cf. Nipoti et al. 2005; K\"{o}rding et al. 2006).

(3) The magnetic field strength estimates along the ridge of the outer double
structure vary slightly from the edges of the lobes towards the centre of the old
cocoon. Unexpectedly, the field strengths in the inner lobes resulting from the
dynamical analysis seem to be twice lower than the corresponding values
estimated in the spectral-ageing analysis. This discrepancy arises from the assumed
different adiabatic indices in the equations of state: a non-relativistic one for the
outer structure and the relativistic one for the inner double structure. The
non-relativistic equation of state implies some energy dissipation process between
magnetic field and thermal particles whose presence in the old cocoon is expected.
Therefore in the non-relativistic conditions for the magnetic energy evolution, a
stronger magnetic field would be necessary to account for the observed brightness
of the inner lobes.

\acknowledgements

The authors are grateful and indebted to Drs. Vicky Safouris and Ravi Subrahmanyan for
access to the archival ATCA and VLA data, to Drs. Matteo Murgia and Karl-Heinz Mack for access to the {\sc SYNAGE} software, and to the anonymous referee for valuable remarks and suggestions that helped us to improve the paper. This project 
was supported in part by the MNiSW funds for scientific research in years 2009--2012 under contract No. 3812/B/H03/2009/36.

\vspace{-10mm}
\begin{appendix}
\section{}
\vspace{-7mm}
\begin{table*}[h]
\caption{Notations for the free parameters in the DYNAGE algorithm and other physical
parameters used through the paper (excluding observational parameters of the radio
source: $D$, $R_{\rm T}$, $V$, $P_{\nu}$, $\alpha_{\nu}$, defined in the
text).}
\begin{tabular*}{148mm}{lcl}
\hline
\hline
Symbol   & Dimension & Parameter\\
\hline
 & {\bf Model} & {\bf free parameters to be assumed}\\
$a_{0}$   & [kpc]      & radius of central core\\
$\beta$   & [dim.less] & exponent of ambient medium-density profile\\
$\gamma_{\rm min}$, $\gamma_{\rm max}$
          & [dim.less] & Lorentz factors of relativistic particles\\
$\Gamma_{\rm jet}$, $\Gamma_{\rm B}$, $\Gamma_{\rm x}$, $\Gamma_{\rm c}$
          & [dim.less] & adiabatic indices of the jet, magnetic field,ambient medium, and cocoon as a whole\\
$\zeta$   & [dim.less] & initial ratio of energy density of the magnetic field to that of particles\\
$k^{\prime}$& [dim.less] & ratio of energy density of thermal particles to that of relativistic ones\\
$\theta$   & [$^{\circ}$] & orientation of the jet's axis\\
\hline
 & {\bf Model} & {\bf parameters to be fitted}\\
$\alpha_{\rm inj}$ & [dim.less] & injection spectral index\\
$t$           & [Myr]  & dynamical age\\
$Q_{\rm jet}$ & [W]    & jet power\\
$\rho_{0}$    & [kg\,m$^{-3}$]& central core density \\
\hline
 & {\bf Other}& {\bf physical parameters}\\
$\beta_{sc.A}$, $\beta_{sc.B}$ & [dim.less] & exponents of ambient density profile in the
self-consistent age solution A and B\\
$p_{\rm hs}$, $p_{\rm min}$, $p_{\rm c}$  & [N\,m$^{-2}$] & hot-spot pressure, its minimum
(equipartition) pressure, and cocoon pressure\\
$u_{\rm e}$, $u_{\rm B}$, $u_{\rm c}$ & [J\,m$^{-3}$] & energy density of relativistic particles,
in magnetic field, and in cocoon as a whole\\
$U_{\rm out}$, $U_{\rm inn}$ &  [J] & total energy radiated from outer and inner lobes\\
$B$, $B_{\rm eqv}$,$B_{\rm iC}$& [nT]& strength of magnetic field, equipartition field, and inverse-Compton field\\
$k$           & [dim.less]    & ratio of energy density of relativistic particles to that of electrons\\
$n_{\rm p}$, $n_{\rm e}$, $n_{\rm g}$ & [m$^{-3}$] & proton, electron number density,
and cold-gas number density\\
$m_{\rm p}$   & [kg]          & proton mass\\
$\rho_{\rm a}$& [kg\,m$^{-3}$] & ambient medium density\\
$\langle t_{\rm i}\rangle$ & [Myr] & mean of $t$ fit for the opposite lobes; i$\Rightarrow$out, inn\\
$\langle Q_{\rm jet,i}\rangle$& [W] & mean of $Q_{\rm jet}$ fit for the opposite lobes\\
$\langle\rho_{0,{\rm i}}\rangle$& [kg\,m$^{-3}$]& mean of $\rho_{0}$ fit for the opposite lobes\\
$\langle\rho_{\rm a,i}\rangle$ & [kg\,m$^{-3}$]& mean of $\rho_{\rm a}$ derived for the opposite lobes\\
$\tau_{\rm syn}$ & [Myr] & spectral (synchrotron) age\\
$\nu_{\rm br}$   & [GHz] & frequency of spectral break\\
\hline
\end{tabular*}
\end{table*}
\end{appendix}

\end{document}